\documentclass[%
reprint,
 amsmath,amssymb,
 aps,
 pra,
]{revtex4-2}

\usepackage{graphicx}
\usepackage{dcolumn}
\usepackage{bm}

\usepackage{hyperref}
\usepackage{dcolumn}
\usepackage{bm}
\usepackage{makecell}
\usepackage{float}

\makeatletter

\renewcommand{\figurename}{Figure}
\renewcommand{\tablename}{Table}

\renewcommand{\fnum@figure}{\textbf{\figurename~\thefigure}}
\renewcommand{\fnum@table}{\textbf{\tablename~\thetable}}

\renewcommand{\@caption@fignum@sep}{: }

\begin{document}

\title{RNA-like Polyelectrolyte in a Viral Capsid: \\ Molecular Dynamics with Explicit Electrostatic Interactions}

\author{Xintong Jiang \textsuperscript{1}}
\author{Colin Denniston \textsuperscript{1}}
\author{Gonca Erdemci-Tandogan\textsuperscript{1, 2, }}
    \email{Corresponding: gerdemci@uwo.ca}

\affiliation{\textsuperscript{1} Department of Physics and Astronomy, University of Western Ontario, London, ON, N6A
3K7, Canada \\ \textsuperscript{2} Department of Medical Biophysics, University of Western Ontario, London, ON, N6A
3K7, Canada} 

\begin{abstract}
The organization of RNA genomes within viral capsids is primarily controlled by electrostatic interactions between the negatively charged genome and positively charged N-terminal domains of coat proteins. In theoretical approaches, these interactions are commonly captured by mean-field models that smooth capsid charge over the inner surface and treat ionic screening as a continuum. However, charges are localized at discrete N-terminal binding sites and ionic screening arises from correlated ion distributions. Here we use molecular dynamics simulations with explicit ions, explicit water, and full Coulomb electrostatics to simulate a linear polyelectrolyte confined within a model capsid bearing discrete N-terminal-like charge sites. We first validate our approach by simulating a polyelectrolyte in bulk solution and demonstrating that persistence length decreases with increasing salt, matching experimental measurements for single-stranded RNA. When confined within a capsid, radial density profiles shift systematically inward from the capsid wall with increasing salt concentration, in agreement with mean-field predictions. By independently varying charge magnitude, binding-site density, and N-terminal protrusion length, we show that total electrostatic coupling governs global organization while geometric details modulate local genome-wall contact and angular genome organization near N-terminals (within the T=3 architecture, linear genome topology, and monovalent salt range studied here). Across all simulations, equilibration times increase sevenfold with salt, revealing kinetic effects inaccessible to equilibrium theory. These results validate continuum approximations for radial organization while revealing deviations arising from discrete molecular details and establishing a framework for future investigations of genome secondary structure, capsid geometry, and assembly kinetics.
\medskip
\medskip

\noindent \textbf{Significance}
Many ssRNA viruses assemble by packaging a negatively charged genome inside a positively charged capsid. Salt concentration and capsid charge can control genome localization, but most models treat solvent implicitly. We present a coarse-grained molecular dynamics framework including an RNA polyelectrolyte, a capsid with discrete N-terminal-like charge sites, explicit monovalent ions, and polarizable water molecules. The model reproduces the salt-dependent persistence length of free ssRNA and captures predicted trends in confined genome density, while revealing quantitative deviations from theoretical predictions due to discrete ion effects and N-terminal sites. This enables systematic tests of how ionic conditions and capsid charge tune genome organization and establishes a validated foundation for incorporating genome secondary structure, capsid details, and assembly kinetics in future studies.

\end{abstract}

\maketitle

\section{\label{sec:intro} Introduction}

Simple RNA viruses encapsulate their genetic material within a protective protein shell called the capsid \cite{Bancroft1970, Bruinsma2006, Zandi2020}. Under many in vitro conditions, this assembly proceeds spontaneously and is driven predominantly by electrostatic interactions between the negatively charged phosphate backbone of the RNA and the positively charged, structurally disordered N-terminal domains of the coat proteins \cite{Sikkema2007, Ni2012, VanDerSchoot2007, Daniel2010, Zlotnick2000, siber2012energies, siber2008nonspecific}. In addition to viral RNAs, capsid proteins can package synthetic polyanions and charged nanoparticles, demonstrating the dominant role of electrostatics \cite{DouglasYoung1998, Hu2008, Cadena-Nava2012}. A quantitative understanding of these electrostatic mechanisms is therefore important for interpreting in vitro assembly experiments and viral packaging measurements.

The spatial organization of the genome within the assembled capsid also depends on the electrostatic interactions, and is impacted by the balance between genome-capsid attraction and the screening effect of dissolved salt. Early in vitro reassembly experiments with CCMV demonstrated that both pH and ionic strength critically control encapsidation efficiency ~\cite{Adolph1976}. Systematic in vitro experiments with CCMV coat proteins showed packaging across genome lengths spanning from 140 to 12,000 nucleotides \cite{Cadena-Nava2012}, demonstrating that electrostatic interactions can drive packaging across diverse RNA sizes. Time-resolved scattering studies further investigated the nonequilibrium assembly dynamics of CCMV~\cite{Chevreuil2018}. Recent assembly studies with MS2 showed that assembly efficiency is controlled by protein-RNA stoichiometry and protein concentration \cite{Garmann2019, Garmann2022, Williams2024MS2}. 

Electrophoresis measurements suggest that viral particles are negatively charged overall \cite{Serwer1995, Serwer1999, Porterfield2010}. Notably, the packaged genomes of many ssRNA viruses carry more negative charge than is needed to neutralize the N-terminals, a phenomenon known as overcharging, quantified as the charge ratio between the genome and the capsid ($|\text{Q}_{\text{RNA}}|/|\text{Q}_{\text{capsid}}|$), with a negative-to-positive charge ratio of approximately 1.6 across many ssRNA viruses \cite{Belyi2006, Hu2008, Ting2011}. The physical origin and robustness of this overcharging remain active topics of investigation.

Mean-field theories offer a framework for predicting how salt concentration controls genome organization. These models treat the capsid as a uniformly charged spherical shell and ion screening as a smooth continuum, and predict that increasing salt concentration shifts RNA distributions inward from the capsid wall \cite{vanderSchoot2005, Ting2011, Gonca2016,  Li2018SCFT}. These models also assume a single bulk dielectric constant for the solvent, whereas water near charged interfaces is known to be substantially dielectrically saturated \cite{fumagalli2018, bonthuis2011}. Further theoretical developments showed that RNA topology, particularly the degree of branching, can significantly affect optimal genome length and radial organization  \cite{Gonca2014, Gonca2016, Gonca2017}.  Li et al. \cite{Li2017} and Dong et al. \cite{Dong2020} found that icosahedrally patterned capsid charge distributions, reflecting discrete N-terminal binding sites, can further modulate these effects.  However, whether these continuum predictions hold quantitatively when discrete ions, explicit solvent, and molecular-level heterogeneity are retained has not been systematically tested through simulations.

Simulation approaches have probed some of these limitations. Monte Carlo simulations of polyelectrolytes inside a viral shell showed that discrete ion effects alter radial distributions beyond continuum screening predictions \cite{angelescu2006monte, angelescu2007packaging}. In parallel, Zhang et al. combined coarse-grained RNA with electrostatic fields computed from the full CCMV capsid structure, revealing that RNA forms a shell near the inner surface with icosahedrally patterned high-density regions driven by the highly positive N-terminal residues \cite{Zhang2004}. More recent molecular dynamics simulations have shown that confined polyelectrolytes adopt bimodal radial distributions driven by self-repulsion \cite{ElSawy2011} and that charge distribution along flexible N-terminal domains significantly affects genome packaging \cite{Safdari2025}. Mattiotti et al. used coarse-grained oxRNA simulations with mean-field capsid representations to study CCMV RNA organization \cite{Mattiotti2024}. Most recently, Li et al. demonstrated spontaneous T=3 capsid assembly around a flexible genome using conformational-switching subunits \cite{Li2025}. Coarse-grained assembly simulations have also studied assembly pathways and capsomer-polyion co-assembly kinetics \cite{Perlmutter2013, Perlmutter2014, Zhang2013, Zhang2013RSC, Zhang2014}. While prior simulations have made significant advances in modeling assembly pathways and genome organization, these studies have relied on either implicit solvent treatment, simplified bead-spring polymer models, effective interaction potentials, or fixed ionic conditions. A systematic study of salt-dependent genome reorganization that simultaneously includes explicit ions, explicit water, discrete capsid binding sites, and Coulomb electrostatics has not been studied.

Here we use coarse-grained molecular dynamics with Coulomb electrostatics, explicit ions, and explicit water to simulate a polyelectrolyte confined within a model capsid bearing discrete N-terminal-like charge sites. The RNA is represented using the MARTINI-2 force field \cite{Uusitalo2017}, which we validate against experimental measurements of the salt-dependent persistence length of single-stranded RNA \cite{Chen2011}. We compare trends seen in simulated radial density profiles to mean-field predictions across salt conditions. We further decompose the role of capsid charge by performing simulations in which charge is either reduced or redistributed across a larger number of binding sites at fixed ionic strength. We also test the sensitivity of genome organization to N-terminal structural representation. Beyond structural organization, we characterize equilibration timescales, revealing kinetic signatures inaccessible to equilibrium theory.

\vspace{-1.75em}
\section{\label{sec:main_method} Methods}

\begin{figure*}[hbt!]
    \centering
    \includegraphics[width=0.8\linewidth]{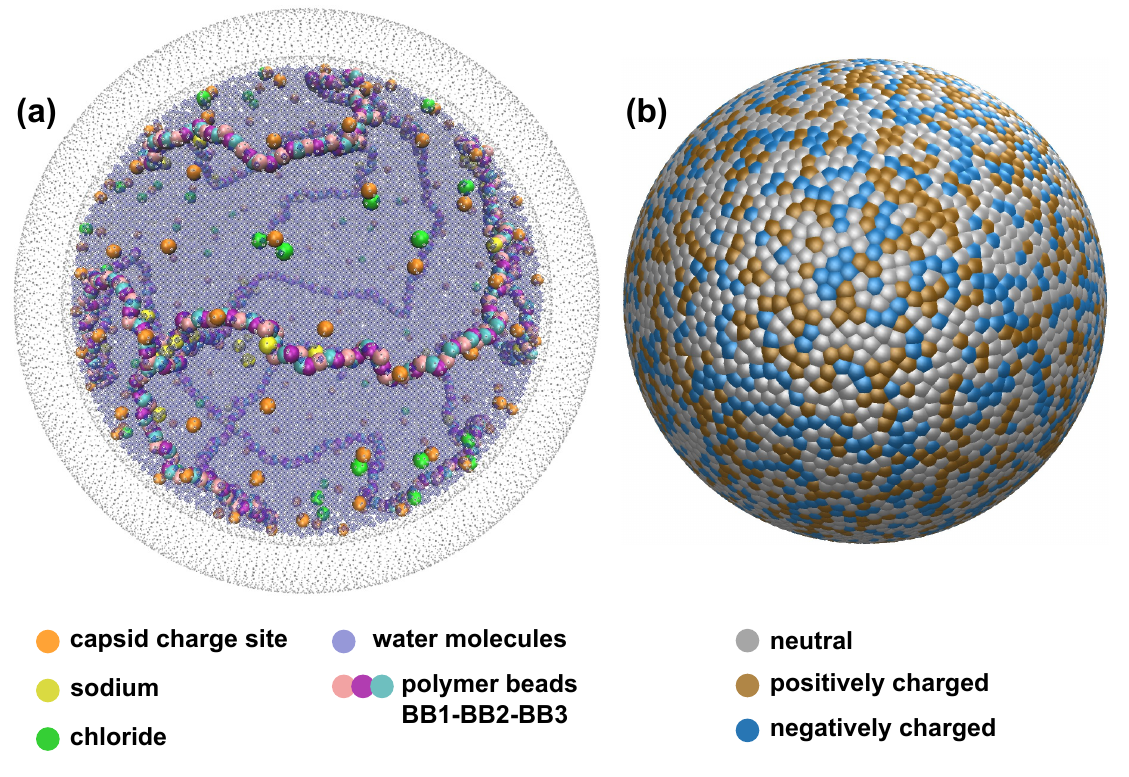}
    \caption{{\bf RNA polyelectrolyte confined within a viral capsid.} (a) The linear polyelectrolyte represents viral RNA using MARTINI-2 backbone beads in a repeating 3-bead pattern: BB1 (pink, phosphate), BB2 (purple, sugar), and BB3 (cyan, sugar). The spherical capsid (gray dots) contains 180 discrete positively charged sites (orange beads) on the inner surface representing N-terminal tails. The aqueous environment includes monovalent ions (sodium in yellow, chloride in green) and polarizable water molecules (purple shaded region). (b) Capsid inner shell composition showing neutral (silver), positively charged (bronze), and negatively charged (blue) beads in a 2:1:1 ratio, yielding an overall neutral shell. Complete model parameters are provided in the SI.}
    \label{fig:Fig1_system_vis}
\end{figure*}

Figure \ref{fig:Fig1_system_vis} illustrates the components of our genome-capsid model. The RNA-like polyelectrolyte is represented as a linear polyelectrolyte chain composed of three distinct bead types, shown in pink, purple, and cyan, respectively (Fig.~\ref{fig:Fig1_system_vis}a, Supplemental Information (SI)). The capsid is modeled as a double-layered spherical shell with inner radius $\sim$12~nm, shown as dotted (gray) spheres. Positively charged binding sites (orange beads, 180 total) are attached to the inner surface to represent N-terminal tails. The aqueous environment includes monovalent ions (sodium in yellow, chloride in green) and polarizable water molecules (light-purple background).

The capsid structure consists of two concentric shells (Fig.~\ref{fig:Fig1_system_vis}b). The outer shell comprises neutral atoms, while the inner shell contains a random 2:1:1 mixture of neutral (silver), weakly positive charged (bronze), and weakly negative charged (blue) beads. This mixed-charge composition prevents artificial ordering of polarizable water molecules at the inner surface \cite{Habibi2014Micelle}. Positive charge for genome binding is provided by 180 discrete sites (orange) attached to the inner surface, representing N-terminal tails.

The polymer backbone uses MARTINI-2 RNA backbone beads \cite{Uusitalo2017} in a repeating 3-bead pattern (BB1, BB2, BB3), but omits the nucleobase and structural restraints of the full MARTINI-2 RNA representation. This sequence-independent chain matches the linear-polyelectrolyte of the continuum theories used for comparison \cite{vanderSchoot2005, Belyi2006, Ting2011, Gonca2014, Gonca2016, Gonca2017, Li2018SCFT}, allowing us to assess the effects of explicit ions, polarizable water, and discrete charge sites without the additional effects from RNA secondary structure. At the same time, we chose not to use a generic bead-spring or freely-jointed polyelectrolyte model, of the kind used in prior simulation studies of confined polyelectrolytes \cite{angelescu2006monte, angelescu2007packaging}. We instead built on the MARTINI-2 RNA framework specifically because it is chemically grounded and extensible: the same backbone parameterization can incorporate base-pairing, establishing this work as a foundation for future studies. The 1122-bead linear polyelectrolyte has a contour length $\sim$336 nm, approximately 14 times the capsid diameter, similar to the ratio used by Williams et al. \cite{Williams2024MS2}. Water is modeled using a coarse-grained three-bead polarizable model \cite{Yesylevskyy2010}, and monovalent ions follow standard MARTINI-2 definitions \cite{Marrink2007Martini}. The capsid is generated by condensing atoms onto a spherical surface \cite{Denniston2022LBfluid}, with 180 positively charged sites attached to the inner surface representing N-terminal tails. The capsid inner radius of $\sim12$ nm (outer radius $\sim$13.6 nm) was chosen to match the capsid size used in prior theoretical work on RNA encapsidation \cite{Gonca2014}, and falls within the physical size range of real ssRNA virus capsids such as CCMV ($\sim$14 nm outer radius) \cite{Cadena-Nava2012, Speir1995CCMV}. The 180 discrete sites correspond to a T=3 architecture by the Caspar-Klug construction ($60\times3=180$ subunits in capsid)\cite{CasparKlug1962}, consistent with the viruses motivating this study (e.g. CCMV, BMV, MS2). Together, these parameters yield a polymer charge of $-374e$ and capsid charge of $\sim$234$e$, giving an overcharging ratio of 1.6 \cite{Belyi2006, Hu2008, Ting2011}. Complete bead definitions and interaction parameters are provided in the Supplemental Information, Sec~\ref{sec:SI_method_beadDef}.

Polymer beads are connected by Finite Extensible Nonlinear Elastic (FENE) bonds \cite{Thompson2022LAMMPS, Kremer1990FENE}
\begin{equation}
\label{eqn:U_FENE}
\begin{split}
U_{\text{FENE}} ={}&
-\frac{1}{2} K R_0^2 \ln
\left[
1 - \left( \frac{r}{R_0} \right)^2
\right] \\
&+ 4\epsilon_{b}
\left[
\left( \frac{\sigma_{b}}{r} \right)^{12}
-
\left( \frac{\sigma_{b}}{r} \right)^6
\right]
+ \epsilon_{b}
\end{split}
\end{equation}
where $K = 30\epsilon_{b}/\sigma_{b}^2$ is the spring constant and $R_0 = 1.5\sigma_{b}$ is the maximum extension \cite{C_Razizadeh_2020, A_Yu_2024}, with $\epsilon_{b}$, the interaction strength of the repulsive Lennard-Jones term between neighboring bonded beads, taken from the corresponding pair interaction (Table~\ref{tab:Supp_MARTNI_bond}) and the $\sigma_{b}$, the length scale of the repulsive Lennard-Jones term between neighboring bonded beads. The value of $\sigma_b$ was chosen so that the minimum of the resulting bonded potential reproduced the equilibrium bond lengths of the MARTINI-2 RNA force field \cite{Uusitalo2017}. The Lennard-Jones (LJ) term in Eq.~\ref{eqn:U_FENE} prevents bonded-bead overlap. Bond-specific parameters are listed in Table~\ref{tab:Supp_MARTNI_bond}.

Non-bonded pair interactions are described by the standard 12-6 Lennard-Jones (LJ) potential
\begin{equation}
    U_{\text{LJ}} = 4\epsilon_{nb} \left[ \left( \frac{\sigma_{nb}}{r} \right)^{12} - \left( \frac{\sigma_{nb}}{r} \right)^6 \right]
    \label{eqn:U_LJ}
\end{equation}
where $r$ is the interparticle distance between the non-bonded particles, $\epsilon_{nb}$ is the well depth, and $\sigma_{nb}$ is the zero-crossing distance. The cutoff is set at the potential minimum: $r_c = 2^{1/6}\sigma_{nb}$. Pair-specific parameters are listed in Table~\ref{tab:Supp_InteractionTab}. Excluded volume is  explicit for every species via a Lennard-Jones potential with a finite core. Na$^+$/Cl$^-$ ions use the standard MARTINI-2 parameterization, which incorporates the first hydration shell into the bead size.

Electrostatic interactions are computed using the Coulomb potential
\begin{equation}
    U_{\text{Coul}} = \frac{C q_i q_j}{\varepsilon r}
    \label{eqn:U_Coul}
\end{equation}
where $q_i$ and $q_j$ are particle charges, $C$ is the energy-conversion constant, and $\varepsilon$ is the dielectric constant. Long-range contributions are computed via the Particle-Particle Particle-Mesh (PPPM) 
method (SI, Sec.~\ref{sec:SI_method_SimProtocol}).

Initial polymer configurations were generated using a random walk algorithm with steric constraints to prevent bead overlap. Polymers were relaxed in bulk solution using harmonic bonds (for 40~ns), which were then replaced with FENE bonds and the dynamics continued for an additional 100~ns, before being compacted and released to generate five independent starting configurations (Fig.~\ref{fig:Vis_DiffConfig}). For capsid simulations, each polymer was placed at the capsid center with explicit water and ions at concentrations of 10, 60, or 100~mM. After energy minimization and charge assignment, production runs used NVE dynamics with a Langevin thermostat at 310~K for 100--600~ns. Complete simulation parameters are provided in the SI, Sec.~\ref{sec:SI_method_SimProtocol}. The three salt conditions, 10, 60, and 100~mM, were chosen to span the range used in prior theoretical work \cite{Gonca2014} that this model is compared against, and are consistent with standard in-vitro assembly buffer conditions across CCMV/MS2 studies \cite{Cadena-Nava2012, ComasGarcia2012, Garmann2014}. Because the genome, ions, and capsid structure occupy physical volume, the realized salt concentration inside the capsid is slightly shifted (by up to $\sim4.5 \%$) from the nominal value used to assign ion counts. We verified this correction empirically from the equilibrated bead counts (Sec.~\ref{sec:SI_IonCorrection}); the correction varies by less than 3\% among the structural variants compared at fixed nominal salt.

In addition to the standard system (180 sites, 1-bead N-terminal), we tested: charge redistributed over 540 and 1470 sites at fixed total capsid charge; 180 sites with total capsid charge reduced (overcharging ratio 2.0); and 0-bead and 2-bead N-terminal representations in place of the standard protrusion (see Discussion for bead length choices).

Persistence length simulations used a shorter 66-bead polymer in bulk solution across varying salt concentrations (10--100~mM). Production runs at 300~K for 400~ns were performed to collect bond-vector trajectories. Bond vectors $\vec{R}_i$ connecting successive beads were used to compute the spatial correlation function $C(j) = \langle \vec{R}_i \cdot \vec{R}_{i+j} \rangle$. Persistence length was extracted by fitting this correlation function to an exponential decay $C(j) = K e^{-j/\lambda}$ and calculating $L_p = \lambda L_B$, where $L_B$ is the mean bond length and $\lambda$ is the number of bonding vectors.

All equilibrium observables were computed from trajectories after the system reached equilibrium. To identify the equilibration time $t_{eq}$, we monitored several properties (e.g., see Fig.~\ref{fig:Full_TimeSeries_Config5}).  The longest quantity to equilibrate was the polymer radius of gyration:
\begin{equation}
    R_g^2 = \frac{1}{M} \sum_i m_i \left( \vec{r}_i - \vec{r}_{\text{cm}}\right)^2
    \label{eqn:Rg}
\end{equation}
where $M$ is the total polymer mass, $m_i$ and $\vec{r}_i$ are the mass and position of bead $i$, and $\vec{r}_{\text{cm}}$ is the center-of-mass of the polymer. For each trajectory, $t_{eq}$ was defined as the earliest time after which $R_g(t)$ remained stable around a stationary mean without systematic drift. Time averages were computed only over the equilibrated interval $t \in [t_{eq}, t_{\text{end}}]$. As an auxiliary check, we also monitored the end-to-end vector $\vec{r}_c = \vec{r}_N - \vec{r}_1$ and its orientation angles (SI Sec.~\ref{sec:SI_method_TrjAnalysis}). Complete time series for all monitored observables for an example set of simulations, originated from the same initial configurations are provided in the SI, Fig.~\ref{fig:Full_TimeSeries_Config5}.

Monomer density as a function of distance from the capsid center was computed using a volume-weighted binning scheme that accounts for finite bead size. Rather than using shells of equal radial thickness, bins were constructed with equal volumes to eliminate geometric artifacts that arise when comparing small inner shells to large outer shells. Each monomer was treated as a hard sphere with effective diameter $d = 2^{1/6}\sigma_{\text{BB-BB}}$, where $\sigma_{\text{BB-BB}}$ is the non-bonded Lennard-Jones interaction distance used in the force-field parameter. When monomers overlapped multiple shells, we calculated what fraction of each sphere belonged to each shell. Details of the weighting scheme and equal-volume binning are provided in the SI, Sec.~\ref{sec:SI_method_rhoCalculation}. Density profiles were normalized by the bin volume and averaged over all snapshots in the equilibrated regime.

Angular organization of the genome around the discrete charge sites was quantified by the angular monomer density. For each charge site, we recorded genome bead centers located within a fixed cone of half-angle $\theta_c$, with apex at the capsid center and axis along the site's radial direction, and binned them into equal-volume angular bins. The angular monomer density profiles were obtained by averaging over all sites and all equilibrated snapshots. Details of angular volume setup and equal-volume binning are provided in SI, Sec.~\ref{sec:SI_AngularProfile}.

All simulations were performed using LAMMPS (Large-scale Atomic/Molecular Massively Parallel Simulator), version 29 Aug, 2024, with the KSPACE, MOLECULE, OPENMP and RIGID package \cite{Thompson2022LAMMPS}. Computational cost across all simulations is reported in SI, Sec.~\ref{sec:SI_ComputationalCost}

\section{\label{sec:results} Results}

We first validated the polyelectrolyte model by measuring persistence length ($L_p$) in free solution across salt  concentrations from 10 to 100~mM. We simulated a 66-bead polymer and extracted $L_p$ from exponential fits to the bond-vector spatial correlation function $C(j) = \langle \vec{R}_i \cdot \vec{R}_{i+j} \rangle$ as described in Methods. An example individual correlation functions and exponential fits is shown in Supplemental Fig.~\ref{fig:Rg_Lp_80NA}.

Figure~\ref{fig:Lp_vs_ion} shows that $L_p$ decreases with increasing salt concentration. At low salt, electrostatic self-repulsion between negatively charged beads stiffens the chain, yielding longer $L_p$. As salt concentration increases, ionic screening weakens self-repulsion, producing a more flexible chain. This is consistent with polyelectrolyte theory. The same trend is captured by the ion-to-polymer charge ratio $Q_{\text{ions}}/Q_{\text{polymer}}$ (Fig.~\ref{fig:Lp_vs_ion} (b)). Our simulated values fall within the experimental bounds reported by Chen et al.~\cite{Chen2011} for 40-nucleotide poly(rU), where $L_p$ decreases from 2.5~nm at low ionic strength to 1.0~nm at high ionic strength, confirming that the MARTINI-based coarse-grained polyelectrolyte model captures salt-dependent flexibility consistent with single-stranded RNA experiments.

\begin{figure}[hbt!]
    \centering
\includegraphics[width=1\linewidth]{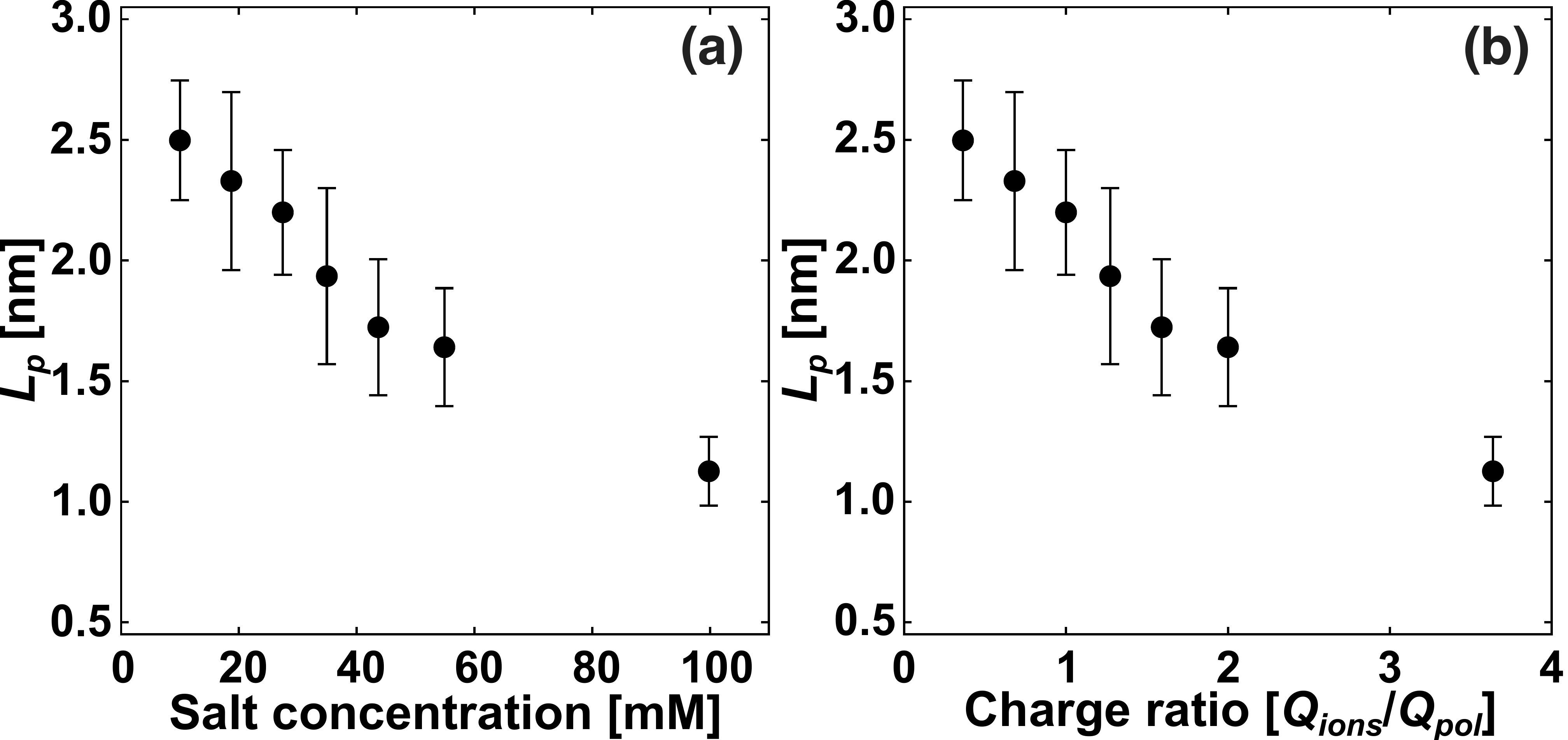}
    \caption{{\bf Ionic screening reduces chain stiffness.} Persistence length ($L_p$) measured in bulk solution as a function of salt concentration (a) and ion-to-polymer charge ratio $Q_{\text{ions}}/Q_{\text{polymer}}$ (b). Error bars are $\pm 1$ standard error of mean over statistically independent samples {(SI, Sec.~\ref{sec:SI_ErrorAnalysis})}.}
    \label{fig:Lp_vs_ion}
\end{figure}

\begin{figure}[hbt!]
    \centering
    \includegraphics[width=0.9\linewidth]{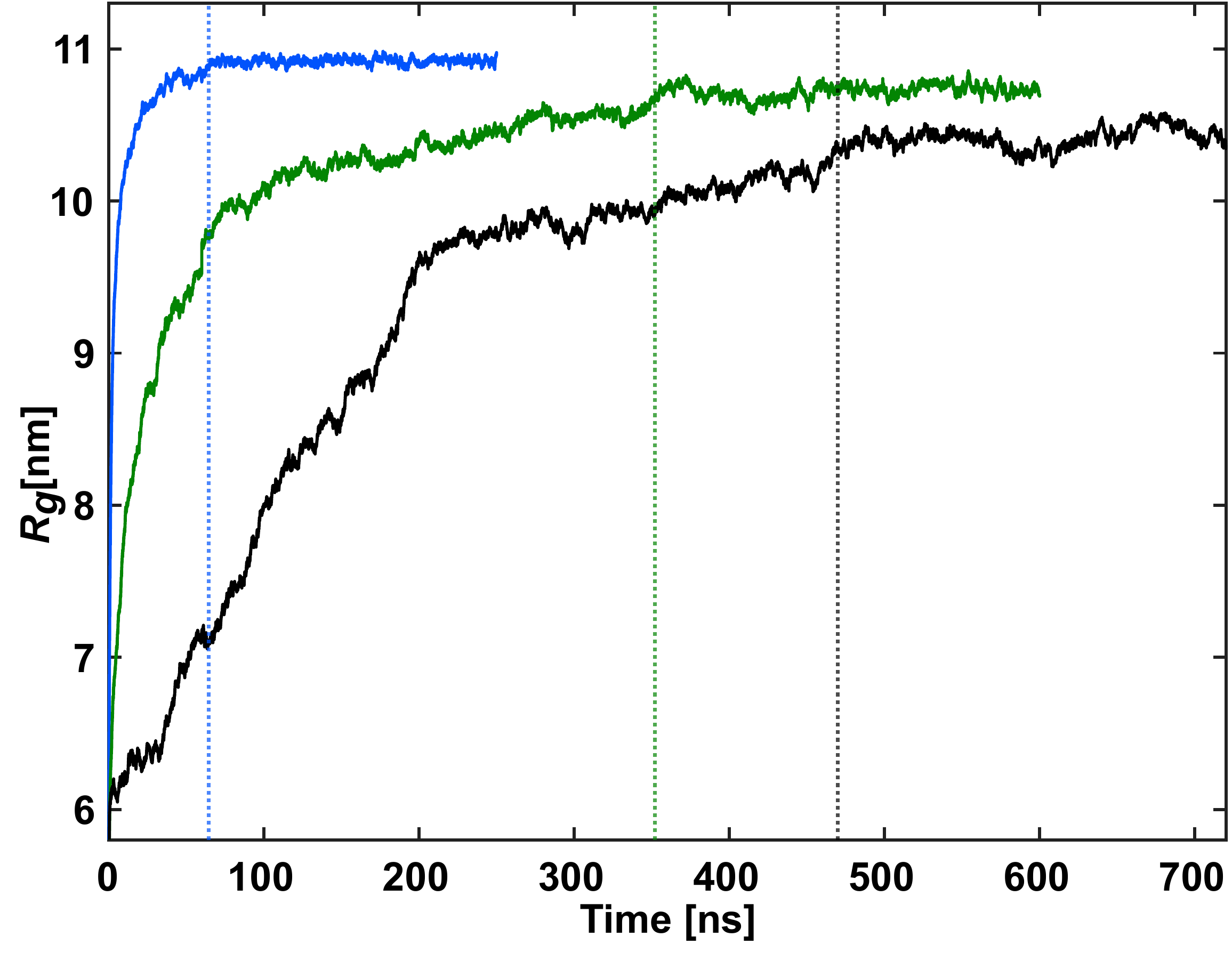}
    \caption{{\bf Polymer relaxation slows with increasing salt.} Representative radius of gyration $R_g(t)$ for three salt concentrations: 10~mM (blue), 60~mM (green), and 100~mM (black), starting from the same initial configuration. Vertical dotted lines mark equilibration times $t_{eq}$: 64~ns (10~mM), 352~ns (60~mM), and 470~ns (100~mM).}
    \label{fig:Rg_Sample}
\end{figure}

Figure~\ref{fig:Rg_Sample} shows representative trajectories of the radius of gyration $R_g(t)$  for simulations at three salt concentrations starting from the same initial configuration (Fig.~\ref{fig:Vis_DiffConfig}e). In all cases, $R_g(t)$ increases from its initial value and plateaus, indicating relaxation to equilibrium. Equilibration times increase systematically with salt: $t_{eq} = 64$~ns (10~mM), 352~ns (60~mM), and 470~ns (100~mM), a trend that holds across all five initial configurations  (SI Fig.~\ref{fig:Individual_Rg}). The slower equilibration at higher salt reflects weaker genome-capsid attraction, which allows the polymer to explore a broader configurational space before settling into equilibrium. All density profiles reported below are computed from equilibrated trajectories ($t \geq t_{eq}$).

\begin{figure}[hbt!]
    \centering
    \includegraphics[width=0.9\linewidth]{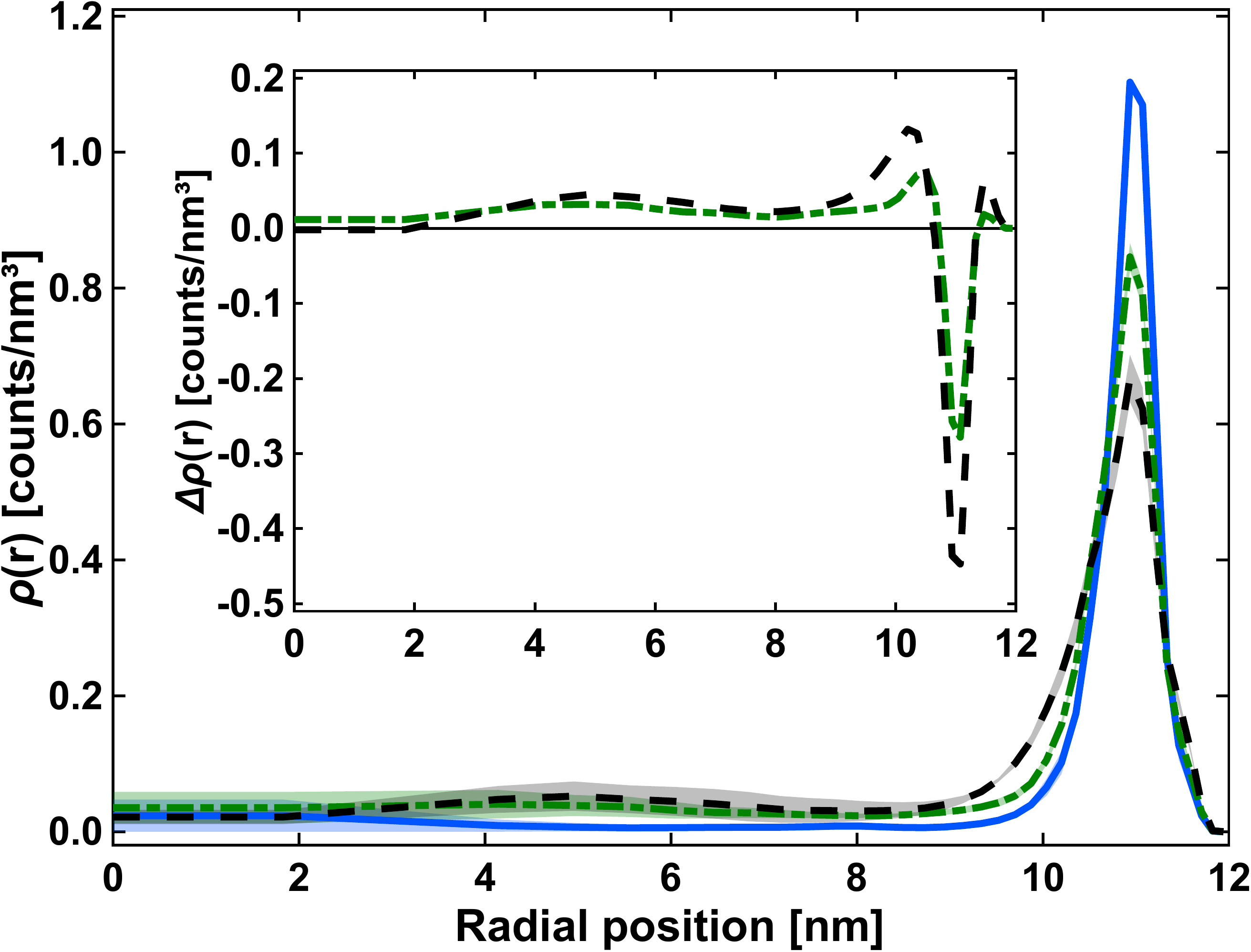}
    \caption{{\bf Increasing salt shifts genome density inward.} Mean radial monomer density $\rho(r)$ versus distance from the capsid center at 10~mM (blue solid), 60~mM (green dot-dashed), and 100~mM (black dashed), each averaged over five independent simulations (shaded bands: standard error of mean across the 5 individual profiles). Inset: Mean density difference $\Delta\rho(r) = \rho(r) - \rho_{\text{10mM}}(r)$ relative to the 10~mM baseline. }
    \label{fig:DensityProfile_Standard}
\end{figure}

 Mean radial density profiles averaged over five independent simulations at each salt concentration are shown in Figure~\ref{fig:DensityProfile_Standard}. Individual profiles for each simulation are in Supplemental Fig.~\ref{fig:Individual_Profiles_Standard}. At 10~mM (Figure~\ref{fig:DensityProfile_Standard}, blue), the profile exhibits a pronounced peak near the capsid wall (r $\approx$ 11 nm), reflecting strong electrostatic attraction between the genome and capsid binding sites. As salt increases to 60 and 100~mM, ionic screening weakens this attraction, allowing entropic effects to compete and drive density inward. The inset quantifies this trend via $\Delta\rho(r) = \rho(r) - \rho_{\text{10mM}}(r)$, confirming a systematic inward shift with increasing ionic screening. The corresponding angular density profiles for the standard capsid at each salt concentration are shown in Supplemental Fig.~\ref{fig:Figure_4_AngularProfile}.

\begin{figure}[hbt!]
    \centering
    \includegraphics[width=0.8\linewidth]{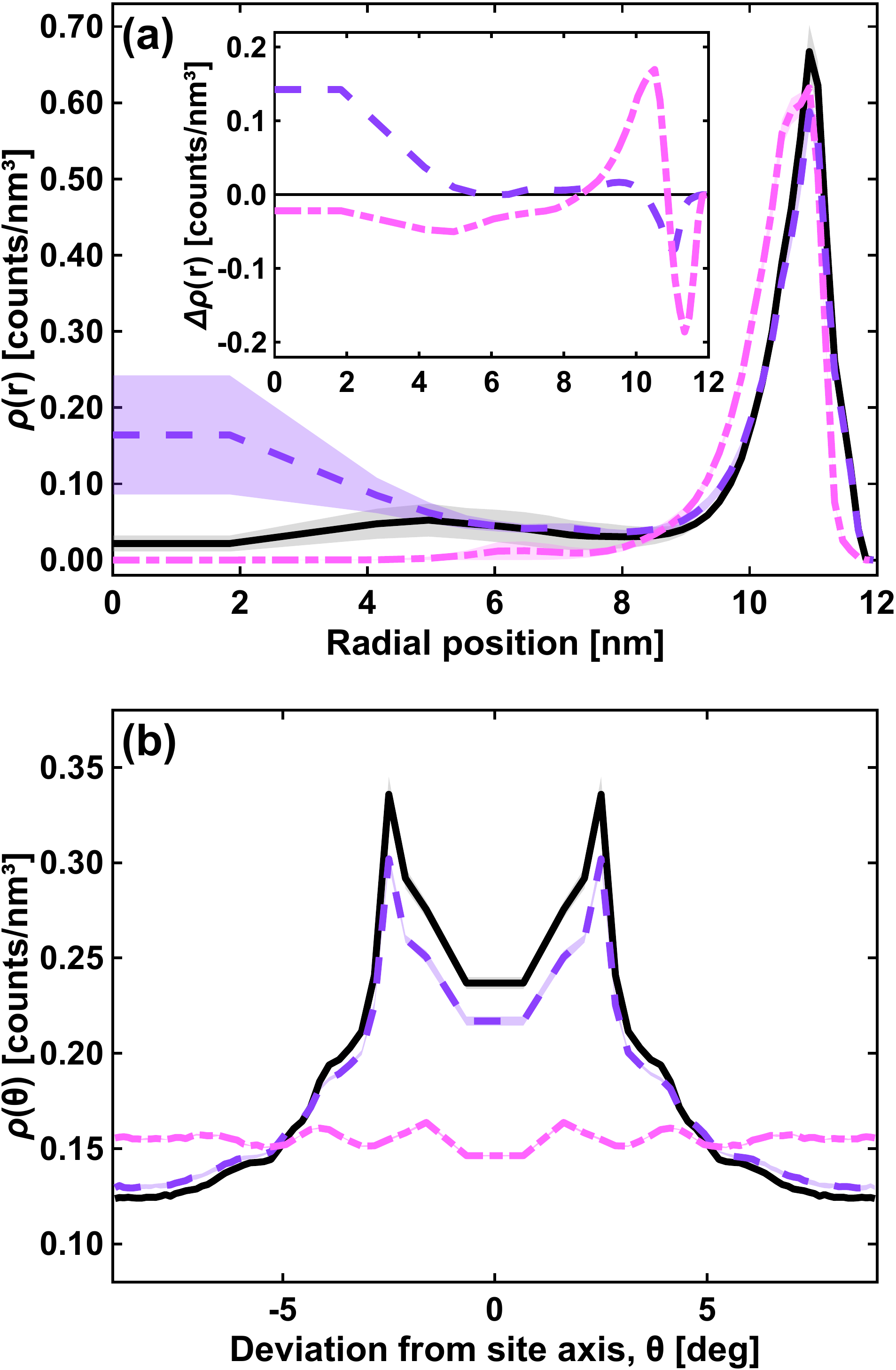}
    \caption{{\bf Charge strength dominates over charge distribution.} (a) Mean radial monomer density $\rho(r)$ at 100~mM salt for three capsid configurations: standard (180 charge sites, overcharging ratio 1.6, black solid), geometrical adjustment (1470 charge sites, overcharging ratio 1.6, pink dot-dashed), and charge reduction (180 sites, overcharging ratio 2.0, purple dashed). Inset: Mean density difference $\Delta\rho(r) = \rho(r) - \rho_{\text{std}}(r)$ relative to the standard profile. (b) Mean angular monomer density profile at 100 mM of salt for standard (black solid), geometrical adjustment (pink dot-dashed), and charge reduction (purple dashed). Shaded bands: standard error of mean across the 5 individual profiles for both panels. }
    \label{fig:DensityProfile_Addition}
\end{figure}

To distinguish geometric effects from electrostatic coupling strength, we performed two additional simulations at 100~mM salt (Fig.~\ref{fig:DensityProfile_Addition}). In the ``geometrical adjustment", we increased the number of discrete charge sites from 180 to 1470 while maintaining the same total capsid charge (+234e, overcharging ratio 1.6). This modification redistributes the capsid charge over more binding sites, reducing site spacing from $\sim$2.9~nm to $\sim$ 1.1 nm, comparable to the free-polymer persistence length at 100~mM (Fig.~\ref{fig:Lp_vs_ion}) and approaching the continuum limit. In the ``charge reduction", we decreased the total capsid charge while maintaining 180 discrete sites, yielding total capsid charge +187.2e and overcharging ratio 2.0, directly weakening the electrostatic coupling strength without changing the geometric distribution of binding sites.

The geometrical adjustment produces only modest changes on the radial density profile: the wall peak decreases slightly and interior density increases weakly (Fig.~\ref{fig:DensityProfile_Addition}(a), pink), with the overall profile shape remaining close to the baseline 100~mM case. In contrast, the charge reduction leads to substantial reorganization (Fig.~\ref{fig:DensityProfile_Addition}(a), purple): wall density decreases markedly while interior density increases, with particularly strong enhancement in the central region ($r < 4$~nm), as reflected in the inset. Some simulations exhibit pronounced central peaks (Supplementary Fig.~\ref{fig:Individual_Profiles_100mMAdditional}, third row), suggesting that weak coupling allows compact, center-localized configurations to become competitive with wall-bound states. Individual profiles are in Supplemental Fig.~\ref{fig:Individual_Profiles_100mMAdditional}. Key parameters are summarized in Table~\ref{tab:Supp_AdditionalRuns}. 

The discrete charge sites break spherical symmetry: the angular profile of the standard capsid (Fig.~\ref{fig:DensityProfile_Addition}b, black) is non-uniform. The genome density is enriched in two peaks flanking each site axis at $\sim \pm 2.5^{\circ}$, separated by an on-axis dip about $30\%$ below the peak maxima. The density varies by a factor of $\sim 2.72$ between its maximum and minimum. Reduced total capsid charge (Fig.~\ref{fig:DensityProfile_Addition}b, purple) at fixed capsid geometry preserves this dual-peak shape at lower amplitude (factor of $\sim2.34$). Redistributing the same total charge over 1470 sites (Fig.~\ref{fig:DensityProfile_Addition}b, pink) removes this shape almost entirely (factor of $\sim 1.12$): the site spacing approaches the continuum limit and the genome resolves individual sites only weakly. The 540-site case (SI Fig.~\ref{fig:Fig_5_4-cases}, factor of $\sim 1.15$) falls between the 180- and 1470-site results, confirming that angular profile heterogeneity decreases with decreasing site spacing.

\begin{figure}[hbt!]
    \centering
    \includegraphics[width=0.8\linewidth]{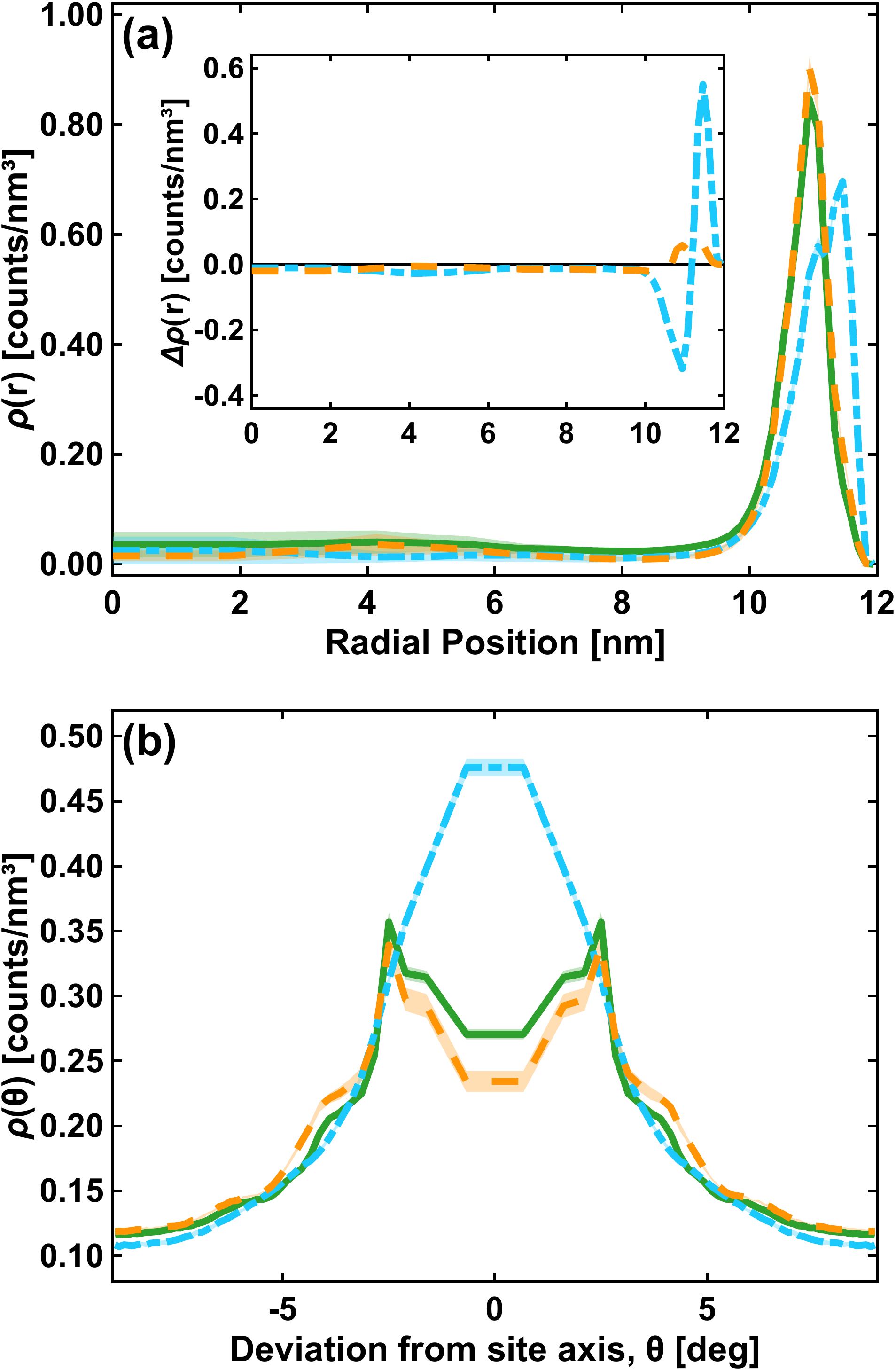}
    \caption{{\bf N-terminal structure affects genome localization.} (a) Mean radial monomer density $\rho(r)$ at 60~mM salt for three N-terminal representations: 0-bead (cyan dot-dashed), 1-bead (green solid), and 2-bead (orange dashed), all with 180 sites and overcharging ratio 1.6. Inset: Mean density difference $\Delta\rho(r) = \rho(r) - \rho_{\text{1-bead}}(r)$ relative to the 1-bead representation. (b) Mean angular monomer density profile at 60~mM salt for 0-bead (cyan dot-dashed),1-bead (green solid), and 2-bead (orange dashed). Shaded bands: standard error of mean across the 5 individual profiles for both panels.} 
    \label{fig:DensityProfile_60mM_Addition}
\end{figure}

To examine how N-terminal representation affects genome distribution, we performed additional simulations at 60~mM testing three structural models while maintaining 180 sites and overcharging ratio 1.6. All simulations reported above use the "1-bead" representation, which models each N-terminal as a discrete charge site protruding from the inner surface. To test this choice, we compared against two alternatives: the "0-bead" representation, which incorporates N-terminal charge directly into the inner shell, creating a smooth interior surface; and the ``2-bead" representation, which models N-terminals as flexible two-bead chains connected by FENE bonds, with charge equally distributed between beads. A visualization of the different representation style for the N-terminals is given in Fig.~\ref{fig:Vis_NterminalRep} in SI and individual profiles are given in Fig.~\ref{fig:Individual_Profiles_60mMAdditional} in SI.

The 2-bead representation produces modest changes relative to the 1-bead case, with radial density shifting slightly toward the wall (Fig.~\ref{fig:DensityProfile_60mM_Addition}a, orange/green). The 0-bead representation differs more substantially (Fig.~\ref{fig:DensityProfile_60mM_Addition}a, cyan): the absence of protruding sites allows closer genome-wall approach, enhancing wall localization. These results demonstrate that discrete protruding binding sites introduce geometric constraints that modulate genome-capsid contact beyond electrostatic attraction alone.

The angular profiles (Fig.~\ref{fig:DensityProfile_60mM_Addition}b) distinguish the three N-terminal representations more sharply than the radial profiles and identify the origin of the dual-peak behavior seen in Fig.~\ref{fig:DensityProfile_Addition}b. With the charge sites embedded in the inner shell (0-bead, Fig.~\ref{fig:DensityProfile_60mM_Addition}b cyan), the profiles have a single maximum, varying by a factor of $\sim4.45$, centered on the site axis, so that the genome density accumulates directly over the opposite charge. Introducing the protruded representation (1-bead, Fig.~\ref{fig:DensityProfile_60mM_Addition}b green) splits the single maximum into dual peaks that vary by a factor of $\sim3.07$, while further increasing the protrusion alters the magnitude of the peaks slightly (factor of $\sim2.86$) but maintains the overall shape of the curve (2-bead, Fig.~\ref{fig:DensityProfile_60mM_Addition}b orange). The dual-peak shape and on-axis depression indicate that the genome tends to wrap around the protruded N-terminal bead rather than stacking directly onto it, consistent with the excluded volume of the bead itself.

\vspace{-1em}
\section{\label{sec:discussion} Discussion}

Explicit molecular dynamics simulations with discrete ions, explicit water, and full electrostatic interactions provide a powerful framework for investigating confined polyelectrolyte dynamics. Rather than imposing salt effects through effective potentials, this approach lets ionic screening emerge naturally from explicit ion dynamics, capturing both equilibrium organization and kinetic pathways. Three key insights emerge from this molecular-level perspective: salt-dependent reorganization occurs through ionic screening of genome-capsid interactions, total electrostatic coupling strength dominates over charge distribution spacing, and discrete protruding binding sites modulate genome localization compared to smooth charged surfaces. Additionally, equilibration times increase substantially with salt concentration, revealing kinetic effects inaccessible to continuum theories.

Salt-dependent genome reorganization follows mean-field predictions \cite{vanderSchoot2005, Ting2011, Gonca2016,  Li2018SCFT}, with density shifting away from the capsid as ionic screening increases. This agreement validates the underlying physical mechanism despite discrete molecular details absent from continuum models. At the same time, we observe systematic deviations in the shape and magnitude of the density profiles, which can be attributed to discrete ion correlations and the finite size of binding sites, effects that are inherently absent from mean-field approaches. Thus, while continuum theories capture the leading-order behavior, our results quantify the contributions that arise when electrostatics is treated at the molecular level. 

The comparison between geometric and electrostatic modifications at fixed salt concentration allows us to disentangle, within the present parameter regime, the relative roles of charge distribution and total coupling strength. If spatial arrangement alone were the dominant factor, redistributing a fixed total charge over a larger number of binding sites would lead to a substantial reorganization of the genome. Instead, we observe only modest changes, indicating that, under these conditions, the precise spacing of binding sites plays a subleading role. In contrast, reducing the total capsid charge produces a pronounced shift of the genome toward the capsid interior, consistent with expectations for weaker coupling regimes \cite{Lee2008}. This behavior reflects that the total electrostatic attraction sets the dominant energetic bias toward the capsid surface, while geometric redistribution primarily perturbs local binding without significantly altering this global balance. 

We emphasize, however, that this conclusion does not preclude an important role for spatial charge localization more generally. Previous studies have shown that discrete and icosahedrally organized N-terminal domains can strongly influence genome organization and optimal packaging \cite{Li2017, Dong2020}. Our results instead suggest that, at fixed salt concentration and within the range of interaction strengths considered here, these geometric effects enter as corrections to a leading-order dependence on total electrostatic coupling. Taken together, this points to a regime-dependent hierarchy in which total charge controls global organization, while spatial distribution modulates finer structural features.

The representation of N-terminal domains introduces additional geometric constraints that are not captured by models in which charge is uniformly distributed over the capsid surface. This 0/1/2-bead series was designed to test the sensitivity of genome organization to N-terminal protrusion length, rather than to represent a specific N-terminal domain. In the absence of explicit protruding binding sites (0-bead model), the genome can approach the capsid wall more closely, leading to enhanced surface localization. Introducing discrete binding sites (1-bead model) imposes a steric constraint that limits this approach, while extending these sites into short flexible chains (2-bead model) produces only modest additional changes. These results suggest that, for short protrusions and at fixed electrostatic conditions, steric accessibility to the capsid surface is the dominant geometric factor influencing genome localization, while further increases in N-terminal length over this narrow range introduce comparatively small corrections.

Recent work by Safdari et al. \cite{Safdari2025} found that extending N-terminals from 1 to 2 beads significantly affects optimal genome length, whereas our simulations show more modest effects on radial organization at fixed chain length. This distinction suggests that N-terminal structure may play a more pronounced role in determining encapsulated mass than in shaping radial density profiles under the conditions considered here. We acknowledge this effect could accumulate further over the longer N-terminal arms not tested here.

In addition to the kinetic behavior noted earlier, two further features are not readily described by continuum theory, or by prior coarse-grained assembly simulations that represent solvent and ionic screening implicitly through effective interaction potentials \cite{Perlmutter2013, Perlmutter2014, Zhang2013, Zhang2014}. The first is the angular genome binding preference revealed by Fig.~\ref{fig:DensityProfile_Addition}b and Fig.~\ref{fig:DensityProfile_60mM_Addition}b, which follows from the discreteness of the binding sites and is substantially suppressed as their spacing approaches the persistence length of the genome. The second is the solvent response: continuum theories assume a single bulk dielectric constant, whereas water near charged interfaces is substantially dielectrically saturated \cite{fumagalli2018, bonthuis2011}, an effect the present polarizable water model captures directly. Consistent with this, explicit-solvent simulations of polyelectrolyte solutions show dynamics qualitatively different from implicit-solvent treatments due to solvent-mediated coupling \cite{Carrillo2023}.

Several limitations point toward future extensions. Here we model the genome as a linear polyelectrolyte without base-pairing or secondary structure; incorporating RNA-specific interactions will further enhance predictions of organization and stiffness \cite{Gonca2014, Gonca2016, Gonca2017, Perlmutter2013}. Salt conditions here are restricted to monovalent ions. Extending to multivalent cations can capture additional electrostatic effects for RNA folding \cite{Chen2011, Chevreuil2018, Bugea2024, Marichal2021, Garmann2014, Cadena-Nava2012, Tresset2017, Saintome2016}. N-terminal models remain simplified compared to longer flexible chain representations \cite{Safdari2025}. Our 1-bead and 2-bead models probe only the shortest protrusion regime, and longer N-terminal chains may introduce additional conformational effects on genome binding that are not captured here. The approach also focuses on equilibrium organization within pre-formed capsids rather than the assembly process itself. 

Typical RNA has secondary structure or branching. The salt-dependent inward shift of genome density and the persistence length validation seen here, which reflect generic polyelectrolyte screening physics, should remain valid in the more general case. However, the relative weighting of total coupling strength against charge distribution is likely to be topology-sensitive. In mean-field theory, base-pairing alone drives a swollen-coil-to-globule transition, and, with increasing RNA length, a first-order transition producing a kinked density profile has been reported \cite{Gonca2017}, so a branched genome could yield different profiles.

The capsid's ionic environment is likewise prescribed. Assembly draws ions into the capsid from a complex bulk solution \cite{Garmann2014}, and the precise ionic content ultimately trapped inside a closed capsid is difficult to determine experimentally. Like the continuum theories, we adopt the same assumption of fixed bulk ionic strength, equilibrating ion content to a specified target concentration (10, 60, or 100 mM). Since screening in the present work emerges from ion dynamics rather than being imposed, the present framework is well positioned to test this assumption directly in future work.

The sevenfold increase in equilibration time with salt suggests that kinetic effects may be significant during assembly. Future extensions to assembly pathways will enable direct comparison to experimental assembly studies \cite{Garmann2019, Garmann2022}. The present explicit-ion, explicit-water framework could also be extended to study genome condensation by freely diffusing RNA-binding domains prior to assembly, as occurs with nucleocapsid proteins in coronaviruses \cite{Li2022SARSCoV2, Zhang2024SARSCoV2}.This would require simulating the charged binding domains and genome directly in solution, without the fixed capsid geometry used here.

\vspace{-1.5em}
\section{\label{sec:conclusion} Conclusion}

Using molecular dynamics with explicit ions, explicit water, and full Coulomb electrostatics, we showed that a MARTINI-based RNA-like polyelectrolyte reproduces the experimentally observed salt-dependent persistence length of ssRNA, and that confined genome density shifts inward with increasing salt, consistent with mean-field predictions. By independently varying charge magnitude, binding-site density, and N-terminal protrusion length, we found that total electrostatic coupling governs global radial organization, while binding-site geometry sets local, angular genome-wall contact. Equilibration times increased sevenfold across the salt range studied, pointing to kinetics not captured by continuum theory.  This study establishes a foundation for systematic exploration of how molecular details govern genome organization in viral capsids. Extending this framework to include RNA secondary structure \cite{Gonca2014, Gonca2016, Gonca2017, Perlmutter2013}, capsid details, capsid size \cite{Wynne1999, Olson1990NBV, Saper2013} and assembly kinetics will work toward a comprehensive and mechanistic picture of RNA virus packaging. 

\vspace{1em}

\vspace{-2.9em}
\section*{Data Availability}
The manuscript and Supplemental Information report complete methodological details and all data. Processed data underlying the figures, analysis code, and simulation input files are available from the authors upon reasonable request.

\vspace{-2em}
\section*{Author Contributions}
XJ, CD, and GE-T designed the research. XJ performed all simulations and analyzed the data, with input from CD and GE-T. XJ and GE-T wrote the manuscript, with input from CD. GE-T and CD acquired funding.

\vspace{-1em}
\section*{Declaration of Interests}
The authors declare no competing interests.

\vspace{-2em}
\section*{Acknowledgments}
We thank the Natural Sciences and Engineering Research Council of Canada (NSERC) for financial support (CD: RGPIN-2025-06298; GE-T: RGPIN-2023-03873) and the Strategic Support for Tri-Council Success Seed Grant of Western University (GE-T). This research has been enabled by the use of computing resources provided by Shared Hierarchical Academic Research Computing Network (SHARCNET) and the Digital Research Alliance of Canada (RRG $\#$5821 [RRG 2026]).

\vspace{-1.5em}
\bigskip
\bibliography{Ref}

\clearpage

\onecolumngrid
\setcounter{page}{1} 
\setcounter{section}{0}
\setcounter{figure}{0}
\setcounter{table}{0}
\setcounter{equation}{0}
\renewcommand{\thesection}{S\arabic{section}}
\renewcommand{\thesubsection}{S\arabic{section}.\arabic{subsection}}
\renewcommand{\p@subsection}{}
\renewcommand{\thetable}{S\arabic{section}.\arabic{table}}
\renewcommand{\thefigure}{S\arabic{section}.\arabic{figure}}
\renewcommand{\theequation}{S\arabic{equation}}
\setlength{\parskip}{1em}

\begin{center}{\fontsize{15pt}{10pt}\bfseries 
RNA-like Polyelectrolyte in a Viral Capsid: \\
Molecular Dynamics with Explicit Electrostatic Interactions\\
\vspace{1em}
Supplemental Information}\\

 \end{center}

\section{Supplemental Methods}
\label{sec: SI_method}

\subsection{Bead Definitions}
\label{sec:SI_method_beadDef}

The polyelectrolyte backbone follows a repeating three-bead MARTINI-2 RNA motif: BB1 (phosphate, charge $-e$, MARTINI Q0 type), BB2 (sugar, neutral, SN0 type), and BB3 (sugar, neutral, SNda type). The 1122-bead linear polyelectrolyte has a contour length $\sim$336 nm, approximately 14 times the capsid diameter.

Water was represented using the three-bead polarizable coarse-grained model of Yesylevskyy et al. \cite{Yesylevskyy2010}, with molecular geometry adapted from TIP3P. Each water molecule consists of a neutral central bead (W; mass 24~amu, charge 0) flanked by two oppositely charged side beads (WP: mass 24~amu, charge $+0.46e$; WM: mass 24~amu, charge $-0.46e$). The bonds are fixed at 0.14~nm, constrained by SHAKE algorithm \cite{Yesylevskyy2010}. The total molecular mass of 72~amu represents four real water molecules. Lennard-Jones interactions are assigned only to the central W bead ($\sigma_{nb} = 0.47$ nm, $\epsilon_{nb} = 4.0$~kJ/mol); the charged side beads contribute exclusively to electrostatics. Monovalent ions follow standard MARTINI-2 definitions \cite{Marrink2007Martini}: Na$^+$ (Qd type, charge $+e$) and Cl$^-$ (Qa type, charge $-e$). 


The capsid was constructed by condensing atoms onto a spherical surface \cite{Denniston2022LBfluid} with an inner radius of approximately 12~nm. It comprises two concentric shells: an outer shell of neutral atoms, and an inner shell containing a 2:1:1 mixture of neutral, positively charged ($+0.005e$), and negatively charged ($-0.005e$) atoms. This composition suppresses artificial ordering of the polarizable water at the inner surface \cite{Habibi2014Micelle} while preserving overall shell neutrality. Additionally, 180 discrete charge sites (CS) representing N-terminal tails are anchored to the inner surface, each carrying a charge of $+1.2986e$, yielding a total capsid charge of $Q_{\text{capsid}} = 180 \times 1.2986e \approx +234e$. Combined with the polymer charge $Q_{\text{polymer}} = -374e$, this produces an overcharging ratio of 1.6 \cite{Belyi2006, Hu2008, Ting2011}. Using the 180 capsid charge sites with 1.6 of overcharged ratio as the standard case, we implemented two additional cases: `geometrical adjustment' and `charge reduction'. The `geometrical adjustment' case increased the number of charge sites from 180 to 540 and 1470, while maintaining the same total capsid charge  ($+234e$, overcharge ratio of 1.6), and the `charge reduction' case decreased the total capsid charge from $+234$e to $+187.2$e, to yield an overcharging ratio of 2, while maintaining the same number of capsid charge sites. Detailed changes for the additional cases relative to the standard case were provided in Table ~\ref{tab:Supp_AdditionalRuns}.

\subsection{Simulation Protocols}
\label{sec:SI_method_SimProtocol}

Initial polymer configurations were generated via a random-walk algorithm. Each monomer bead was placed 0.35~nm from the preceding bead with a random angular deviation of $\pm\theta$ relative to the prior bond direction; trial positions were rejected if within 0.47~nm of the second-previous bead to prevent non-adjacent monomer overlap. Polymers were relaxed using harmonic bonds in a $40^3$~nm$^3$ box under NVE dynamics with a Langevin thermostat at 310~K (damping constant 100~fs, timestep 10~fs) for 40~ns, after which harmonic bonds were replaced with FENE bonds and equilibration continued for an additional 100~ns. A condensing-and-releasing procedure was then applied to generate compact conformations compatible with capsid encapsulation. During the condensing stage, a central particle exerting strong Lennard-Jones attractive interactions drew monomers toward the box center. Upon removal of the central particle, the polymer was allowed to diffuse freely for 10--50~ns (releasing stage). Five independent starting configurations were generated (Fig.~\ref{fig:Vis_DiffConfig}).

For capsid--genome production runs, each polymer was placed at the capsid center. Na$^+$ and Cl$^-$ ions were inserted randomly to achieve bulk concentrations of 10, 60, or 100~mM (Table~\ref{tab:Supp_IonConcentration}). Water molecules were placed on a 0.535~nm cubic lattice. A 1~ns soft-core equilibration ramped non-bonded interaction strengths from 0 to 40~kcal/mol to eliminate atomic overlaps, followed by energy minimization to a force tolerance of $10^{-8}$~kcal~mol$^{-1}$~\AA$^{-1}$. Partial charges were assigned as follows: BB1~$= -1e$, BB2~$=$~BB3~$= 0$; WP~$= +0.46e$, WM~$= -0.46e$, W~$= 0$; Na$^+$~$= +1e$, Cl$^-$~$= -1e$; capsid surface patches~$= \pm0.005e$; capsid charge sites~$= +1.2986e$ (standard),$+0.433e$ (geometrical adjustment, 540 CS), $+0.159e$ (geometrical adjustment, 1470 CS), or $+1.04e$ (charge reduction). 

Production runs employed NVE dynamics with a Langevin thermostat at 310~K (damping constant 100~fs, timestep 10~fs). Capsid walls and charge sites were held fixed throughout. Water geometry was constrained via SHAKE with a harmonic angle potential (force constant 0.5~kcal~mol$^{-1}$~\AA$^{-2}$, equilibrium angle $0^\circ$). FENE bonds \cite{Thompson2022LAMMPS, Kremer1990FENE} were assigned to BB1--BB2, BB2--BB3, and BB3--BB1 pairs using parameters $K = 30\epsilon_b/\sigma_b^2$ and $R_0 = 1.5\sigma_b$ \cite{C_Razizadeh_2020, A_Yu_2024}, where $\epsilon_b$, the interaction strength of the repulsive Lennard-Jones term between neighboring bonded beads, was taken from the corresponding non-bonded pair interaction (Table~\ref{tab:Supp_InteractionTab}) and $\sigma_b$ was the length-scale of the repulsive Lennard-Jones term between neighboring bonded beads. The values for $\sigma_{b}$ were chosen such that the resultant equilibrium bond lengths in FENE bond matched the equilibrium bond length in MARTINI RNA model \cite{Uusitalo2017}. Non-bonded interactions combined a Lennard-Jones term (cutoff at the minimum position of corresponding Lennard-Jones potential, $R_{min}$, where $R_{min} = 2^{1/6}\sigma_{nb}$), with long-range Coulomb electrostatics computed via PPPM (dielectric constant 2.5, accuracy $10^{-4}$, real space cutoff 1.2~nm) \cite{Yesylevskyy2010}.  Intramolecular water interactions were excluded to prevent the central W bead from interacting with its own WP and WM beads, and the Lennard-Jones interactions were only granted for the central W bead for all water molecules\cite{Yesylevskyy2010}. Depending on the degree of equilibration required, production runs spanned 100--600~ns, with configurations saved every 12.5~ns (1{,}250{,}000 timesteps) and snapshots saved every 25~ps (2{,}500 timesteps).

Persistence length validation used a shorter 66-monomer polymer simulated in bulk solution across salt concentrations of 10--100~mM. This reduced chain length was chosen to keep system size tractable, as the full 1122-monomer polymer would require a prohibitively large simulation box. The 66-monomer system used identical interaction, bonding, and preparation protocols, with the polymer placed in a $11^3$~nm$^3$ box containing water and ions and soft-equilibrated for 1~ns. Production runs were conducted at 300~K for 400~ns, with snapshots saved every 25~ps (2{,}500 timesteps).

Complete non-bonded interaction and bonding parameters are provided in Tables \ref{tab:Supp_MARTNI_bond}--~\ref{tab:Supp_InteractionTab}. Ion counts for all salt concentrations are listed in Table~\ref{tab:Supp_IonConcentration}. Conditions specific to the geometrical adjustment and charge reduction runs are summarized in Table~\ref{tab:Supp_AdditionalRuns}.

\begin{table}[hbt!]
    \centering
    \begin{tabular}{|c|c|c|c|c|}
        \hline
        \textbf{Beads} & \makecell{\textbf{Bond type} \\ \textbf{(FENE bond)}} & \makecell{\textbf{Equilibrium bond length} \\ \textbf{(MARTINI)}} & \makecell{\textbf{$\epsilon_b$} \\ \textbf{(FENE)}}  & \makecell{\textbf{$\sigma_b$} \\ \textbf{(FENE)}}\\
        \hline
        BB1-BB2 & 1 & 0.363 nm & 3.5 $\text{kJ/mol}$ & 0.375 nm \\
        \hline
        BB2-BB3 & 2 & 0.202 nm & 2.625 $\text{kJ/mol}$ & 0.208 nm \\
        \hline
        BB3-BB1 & 3 & 0.354 nm & 4.0 $\text{kJ/mol}$ & 0.366 nm \\
        \hline
    \end{tabular}
    \caption{\textbf{FENE bond parameters for polymer bead pairs.}
    Here, the $\epsilon_b$ is the interaction strength of the repulsive Lennard-Jones term in FENE potential, which are taken from the corresponding non-bonded pair interactions (Table.~\ref{tab:Supp_InteractionTab}), and $\sigma_b$ is the length-scale of the repulsive Lennard-Jones term in FENE potential. Note that, the values of $\sigma_b$ were chosen such that the equilibrium bond length of the FENE bond matched the equilibrium bond length in MARTINI RNA model \cite{Uusitalo2017}.
    }
    \label{tab:Supp_MARTNI_bond}
\end{table}

\begin{table}[hbt!]
\centering
\begin{tabular}{|c|c|c|c|c|c|c|}
\hline
Particles & \textbf{BB1 (Q0)} & \textbf{BB2 (SN0)} & \textbf{BB3 (SNda)} & $\textbf{Na}^+$ \textbf{(Qd)} & $\textbf{Cl}^-$ \textbf{(Qa)} & \textbf{Water (W)} \\
\hline
\makecell{\textbf{BB1 (Q0)} \\ $q=-e$} & IV & IV & III & II & II & II \\
\hline
\makecell{\textbf{BB2 (SN0)} \\ $q=0$} & -- & IV (75\%, s) & IV (75\%, s) & IV & IV & IV (95\%) \\
\hline
\makecell{\textbf{BB3 (SNda)} \\ $q=0$} & -- & -- & III (75\%, s) & I & I & IV (95\%) \\
\hline
\makecell{$\textbf{Na}^+$ \textbf{(Qd)} \\ $q=e$} & -- & -- & -- & I & O & I \\
\hline
\makecell{$\textbf{Cl}^-$ \textbf{(Qa)} \\ $q=-e$} & -- & -- & -- & -- & I & I \\
\hline
\makecell{\textbf{Water (W)} \\ $q=0$} & -- & -- & -- & -- & -- & III \\
\hline
\makecell{\textbf{Outer capsid} \\ $q=0$} & IV & IV & IV & IV & IV & IV \\
\hline
\makecell{\textbf{Inner capsid} \\ $q=0, \pm0.005e$} & IV & IV & IV & IV & IV & IV \\
\hline
\makecell{\textbf{Charge sites} \\ $q=+1.2986e$ \\ (Standard)} & IV & IV & IV & IV & IV & IV \\
\hline
\end{tabular}
\caption{\textbf{Summary of non-bonded Lennard-Jones interaction parameters.} Interaction levels and notations are consistent with MARTINI force-field in Ref.~\cite{Marrink2007Martini, Uusitalo2017, Yesylevskyy2010}. The interactions are categorized in levels (O-IV). The LJ parameter $\sigma_{nb} = 0.47\text{ nm}$ is applied to all the interactions shown here without the additional note of `s'. In the case of `s', the LJ parameter $\sigma_{nb} = 0.43\text{ nm}$ is applied. Level O: $\epsilon_{nb} = 5.6\ \text{kJ/mol}$. Level I: $\epsilon_{nb} = 5.0\ \text{kJ/mol}$. Level II: $\epsilon_{nb} = 4.5\ \text{kJ/mol}$. Level III: $\epsilon_{nb} = 4.0\ \text{kJ/mol}$. Level IV: $\epsilon_{nb} = 3.5\ \text{kJ/mol}$. The note of percentage shown in this table indicated the scaling of the standard interaction strength. For example, IV (95\%) indicates the interaction strength of $\epsilon_{nb} = 3.5 \times 0.95 = 3.325\ \text{kJ/mol}$.  As the side beads in coarse-grained water molecule (WP and WM beads) do not interact through Lennard-Jones interaction, they are excluded from this table.}
\label{tab:Supp_InteractionTab}
\end{table}

\begin{table}[hbt!]
\centering
\begin{tabular}{|c|c|c|c|c|}
\hline
\textbf{\makecell{Simulation \\ type}} & \textbf{\makecell{Genome charge \\ ($Q_{pol}$, in $e^-$)}} & \textbf{\makecell{Capsid charge \\ ($Q_{cap}$, in $e^-$)}} & \textbf{\makecell{Overcharged \\ ratio \\ ($|Q_{pol}|/|Q_{cap}|$)}} & \textbf{\makecell{No. of \\ N-terminal sites}} \\ \hline
\makecell{Standard \\ (10 mM, 60 mM, 100 mM)} & -374 & 234 & 1.6 & 180 \\ \hline
\makecell{Geometrical adjustment \\ (100 mM)} & -374 & 234 & 1.6 & 540 \\
\hline
\makecell{Geometrical adjustment \\ (100 mM)} & -374 & 234 & 1.6 & 1470 \\ \hline
\makecell{Charge reduction \\ (100 mM)} & -374 & 187.2 & 2 & 180 \\ \hline
\end{tabular}
\caption{\textbf{Summary of key parameters for standard and additional runs.} Here, the standard run refers to the capsid condition of 180 capsid charge sites (each carrying a charge of $+1.2986e$) with overcharging ratio of 1.6. The additional runs refer to the capsid condition adjustment cases of `geometrical adjustment' and `charge reduction'. The `geometrical adjustment' case refers to the capsid condition of which maintained the same total capsid charge as the standard case, but redistributed the charges onto more capsid charge sites (each carrying a charge of $+0.433e$ or $+0.159e$). The `charge reduction' refers to the capsid condition of which maintained the same number of capsid charge sites as the standard case, but decreased the total capsid charge with each site carrying a charge of $+1.04e$, resulting in an overcharging ratio of 2.}
\label{tab:Supp_AdditionalRuns}
\end{table}

\begin{table}[h!]
    \centering
    
    \label{tab:1122LinearPOL_salt}
    \begin{tabular}{|c|c|c|c|}
        \hline
         & \textbf{$\text{Na}^+$} & \textbf{$\text{Cl}^-$} & \textbf{$|{Q}_{\text{Na}}|/|{Q}_{{pol}}|$} \\ \hline
        \textbf{10 mM} & 44 & 44 & 0.118 \\ \hline
        \textbf{60 mM} & 262 & 262 & 0.7 \\ \hline
        \textbf{100 mM} & 436 & 436 & 1.166 \\ \hline
        
    \end{tabular}
    \caption{\textbf{Summary of salt concentrations with the corresponding number of sodium and chloride ions used in the simulations.} The number of ions was set to reproduce the target salt concentration within the internal volume of the capsid.}
    \label{tab:Supp_IonConcentration}
\end{table}

\subsection{Trajectory Analysis}
\label{sec:SI_method_TrjAnalysis}

For a polymer of $N$ monomers, bond vectors $\vec{R}_i$ connect monomer $i$ to monomer $i+1$. The spatial correlation function between bond vectors $\vec{R}_i$ and $\vec{R}_{i+j}$, separated by $j$ bonds along the chain, is
\begin{equation}
    C(j) = \left\langle \vec{R}_i \cdot \vec{R}_{i+j} \right\rangle, \quad j \in [1, N-2]
\label{eq:lp_fitting}
\end{equation}
where brackets denote averaging over all bond pairs and simulation time. The correlation function decays exponentially, and we fit $C(j) = K e^{-j/\lambda}$ to extract the decay length $\lambda$ in units of bond segments. The persistence length in physical units is then $L_p = \lambda L_B$, where $L_B = 0.304$~nm is the mean bond length averaged over BB1-BB2, BB2-BB3, and BB3-BB1 bonds. An example fit is shown in Fig.~\ref{fig:Rg_Lp_80NA}.

\begin{figure}[hbt!]
    \centering
    \includegraphics[width=0.9\linewidth]{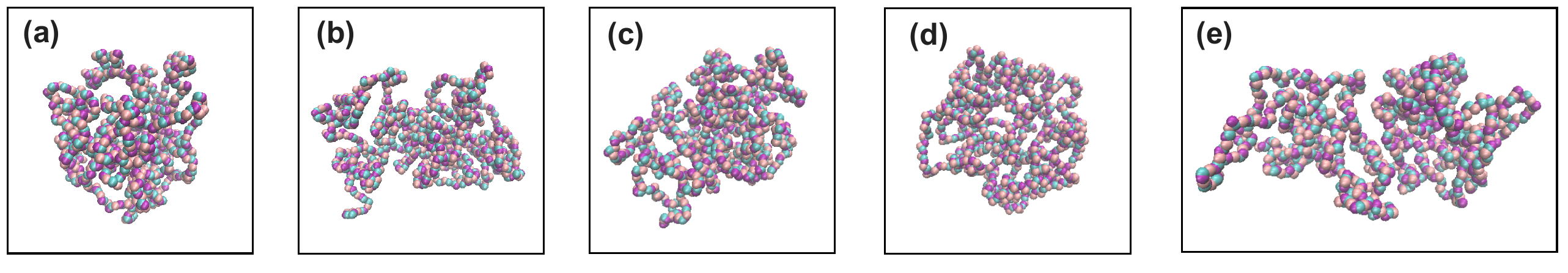}
    \caption{\textbf{Five independent initial polymer configurations.} Panels \textbf{(a)}--\textbf{(e)} show the compact starting conformations used in capsid simulations, generated by the compaction and release procedure described in Methods. Pink, purple, and cyan beads represent BB1, BB2, and BB3 monomers, respectively.}
    \label{fig:Vis_DiffConfig}
\end{figure}

\begin{figure}[H]
    \centering
    \includegraphics[width=0.7\linewidth]{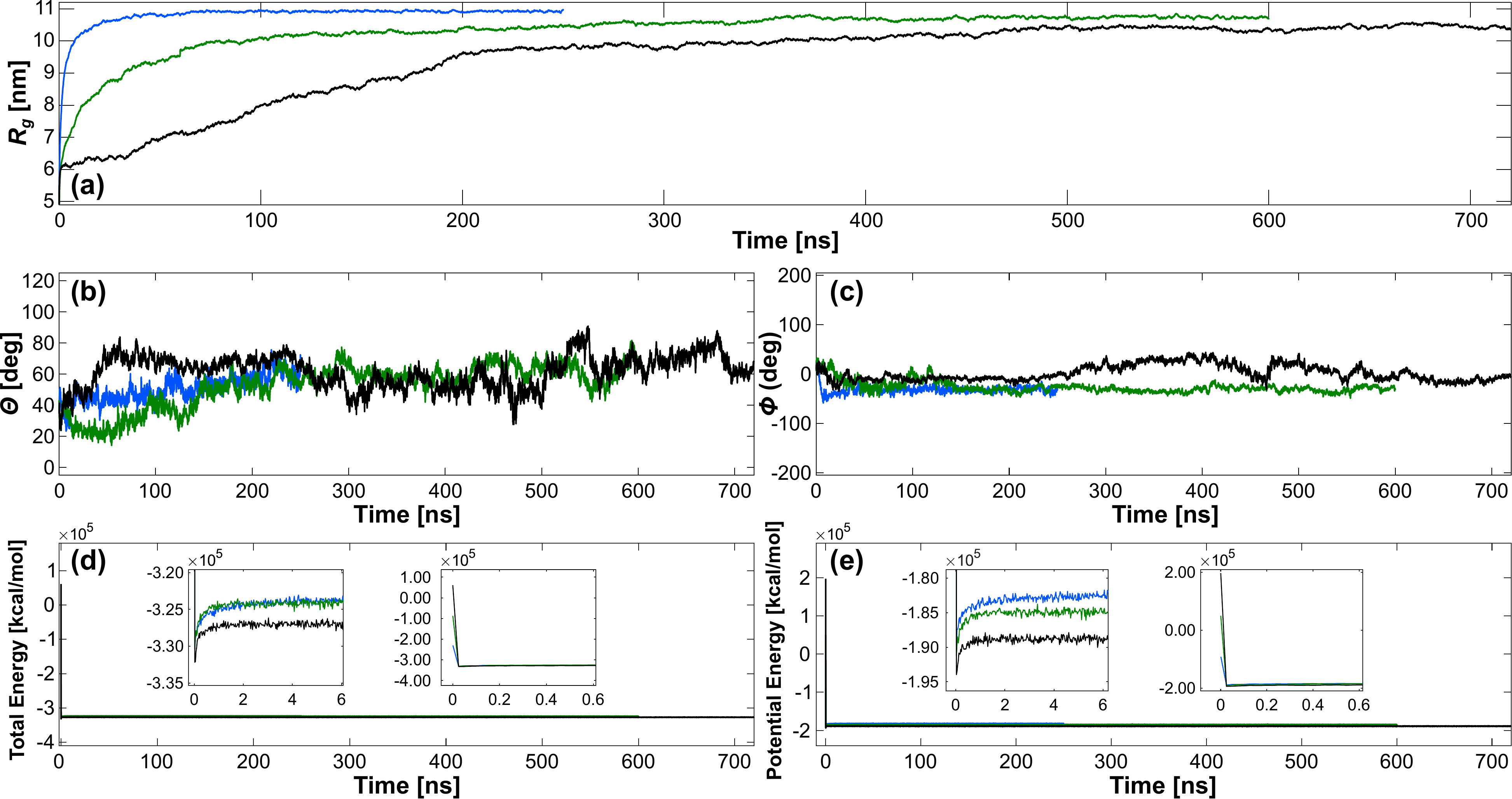}
    \caption{\textbf{Time series of equilibration observables for simulations starting from initial configuration 5 (Fig.~\ref{fig:Vis_DiffConfig}e) at 3 different salt concentrations.} \textbf{(a)} Radius of gyration $R_g(t)$. \textbf{(b)} Polar angle $\theta(t)$ and \textbf{(c)} azimuthal angle $\phi(t)$ of the end-to-end vector (Eq.~\ref{eq:end2end_orientation}). \textbf{(d)} Total energy and \textbf{(e)} potential energy of the system. In \textbf{(d)} and \textbf{(e)}, insets show the first 6~ns and 0.6~ns respectively, highlighting early transient behavior. In each panel, curves correspond to: 10~mM standard (blue), 60~mM standard (green), and 100~mM standard (black). Among all monitored observables, $R_g(t)$ exhibits the slowest relaxation and was therefore chosen as the equilibration criterion.}
    \label{fig:Full_TimeSeries_Config5}
\end{figure}

\begin{figure}[H]
    \centering
    \includegraphics[width=0.8\linewidth]{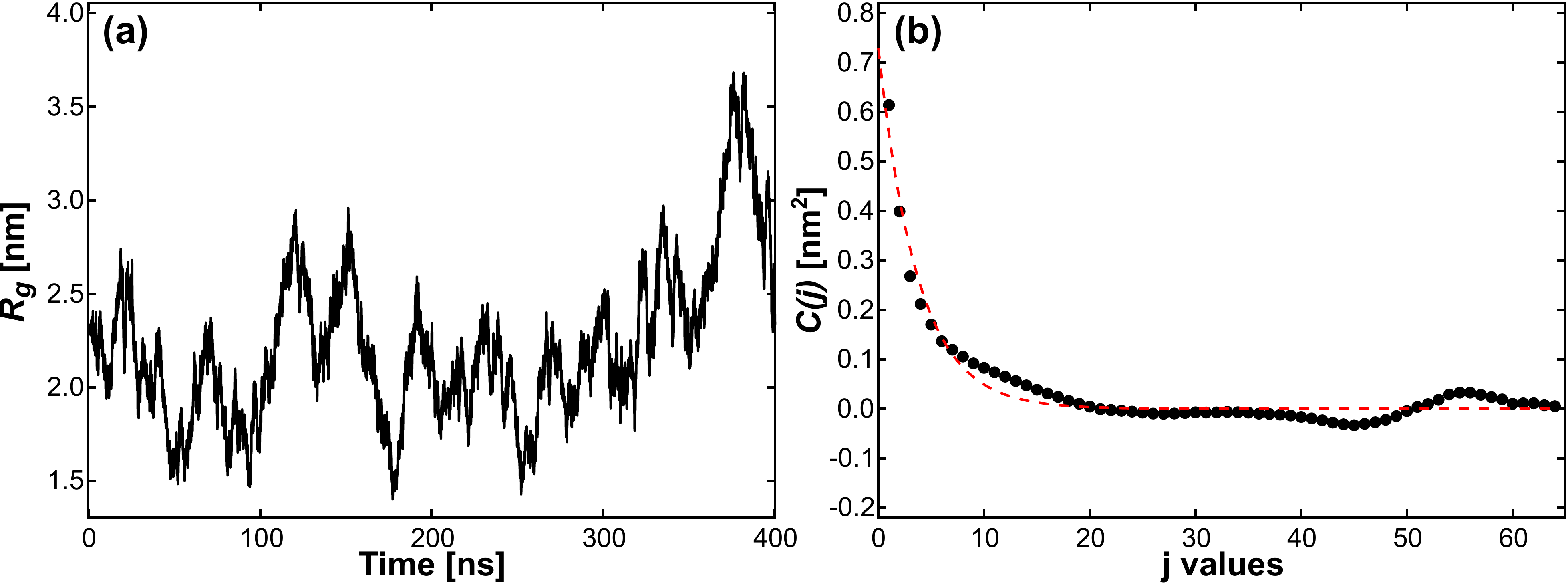}
    \caption{\textbf{Persistence length extraction for the 66-monomer polymer in free solution at 100~mM salt.} \textbf{(a)} Radius of gyration $R_g(t)$. \textbf{(b)} Bond vector correlation function $C(j)$ (black dots) and exponential fit (red dashed line), as defined in Eq.~\ref{eq:lp_fitting}. The fit was performed over the first 10 data points, yielding $\lambda \approx 3.70$ bond segments. Persistence length simulations used a shorter 66-monomer polymer in free solution rather than the full 1122-bead capsid polymer to reduce computational cost.}
    \label{fig:Rg_Lp_80NA}
\end{figure}

\begin{figure}[H]
    \centering
    \includegraphics[width=1\linewidth]{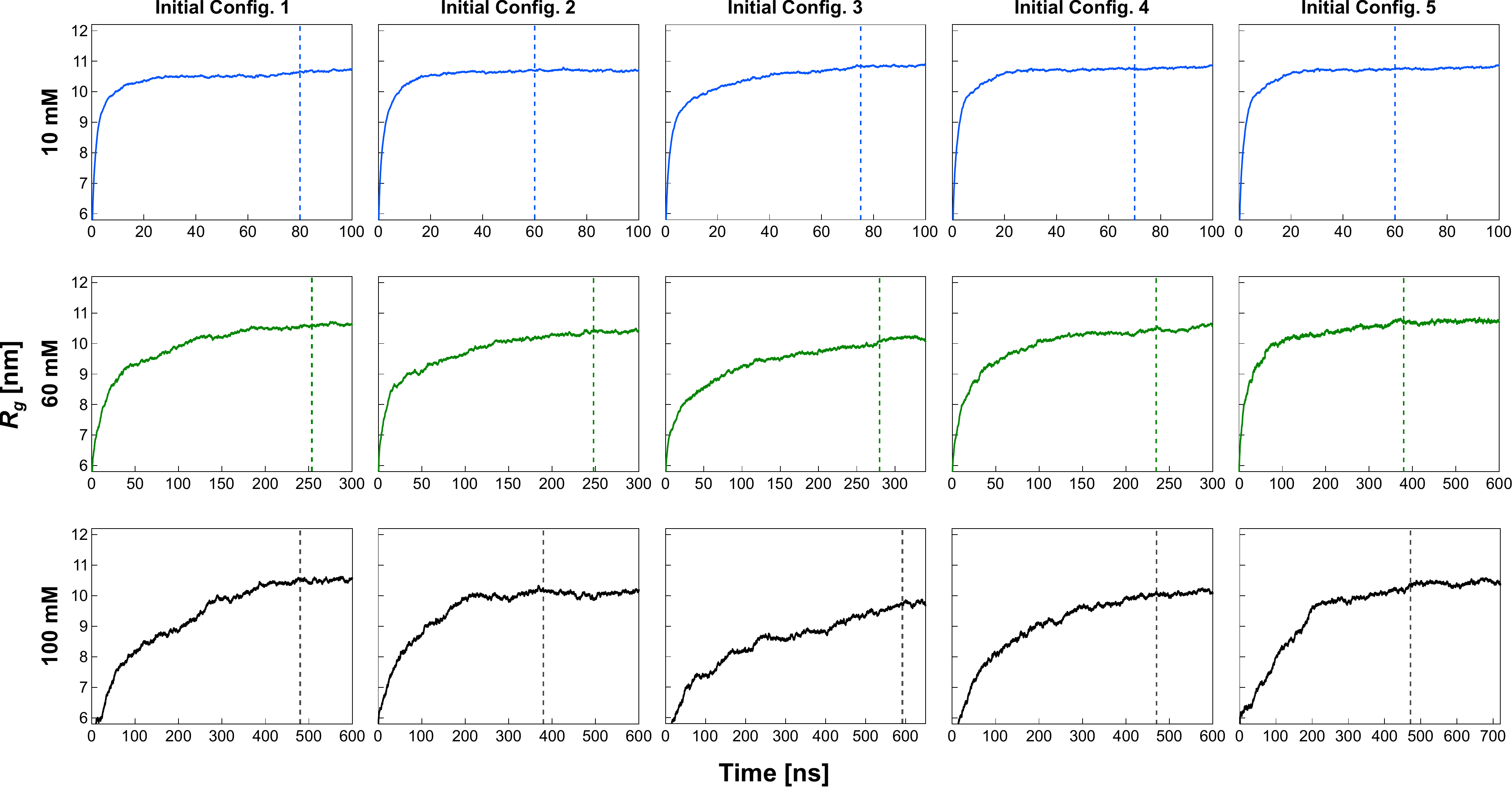}
    \caption{\textbf{Radius of gyration $R_g(t)$ for all five initial configurations at three salt concentrations.} For each row, from left to right, plots correspond to initial configurations 1--5 (Fig.~\ref{fig:Vis_DiffConfig}a--e), respectively. Curves show 10~mM (blue), 60~mM (green), and 100~mM (black). Vertical dashed lines mark the equilibration time $t_{eq}$ for each simulation.}
    \label{fig:Individual_Rg}
\end{figure}

Equilibration was assessed primarily through the radius of gyration $R_g(t)$, as described in the main text. As an auxiliary check, we also monitored the end-to-end vector orientation via polar and azimuthal angles:
\begin{equation}
    \theta = \arccos(u_{c,z}), \quad \phi = \arctan2(u_{c,y}, u_{c,x})
\label{eq:end2end_orientation}
\end{equation}
where $\vec{u}_c = \vec{r}_c/|\vec{r}_c|$ is the unit end-to-end vector. End-to-end reorientation corresponds to the slowest relaxation mode of the polymer; stable $\theta(t)$ and $\phi(t)$ confirm that the system has reached equilibrium. Representative time series for $R_g(t)$, $\theta(t)$, $\phi(t)$, and energies for initial configuration 5 (Fig.~\ref{fig:Vis_DiffConfig}e) are provided in Fig.~\ref{fig:Full_TimeSeries_Config5}.

\subsection{Radial Density Profile Calculation}
\label{sec:SI_method_rhoCalculation}

Each monomer is treated as a hard sphere with diameter $d = 2^{1/6}\sigma_{\text{BB-BB}}$ and radius $R_m = d/2$. To account for finite bead size, the contribution of each monomer to a radial bin is weighted by the fraction of its volume that falls within that bin.

For a monomer centered at distance $r_0$ from the capsid center, the volume of the monomer lying within a sphere of radius $R$ is:
\begin{equation}
V_{\text{in}}(R,r_0;R_m) = 
\begin{cases}
0, & r_0 \ge R + R_m \\
\frac{4\pi}{3}R_m^3, & r_0 \le R - R_m \\
\frac{\pi(R+R_m-r_0)^2}{12r_0} P(R,r_0,R_m), & |R-R_m| < r_0 < R+R_m
\end{cases}
\end{equation}
where $P(R,r_0,R_m) = r_0^2 + 2r_0 R_m - 3R_m^2 + 2r_0 R + 6R_m R - 3R^2$. The weighted contribution of monomer $j$ to bin $i$ spanning $[R_i, R_{i+1})$ is:
\begin{equation}
    w_{ij} = \frac{V_{\text{in}}(R_{i+1}) - V_{\text{in}}(R_i)}{V_m}
\end{equation}
where $V_m = (4/3)\pi R_m^3$ is the monomer volume. Only monomers whose overlap range $[r_0 - R_m, r_0 + R_m]$ intersects a bin boundary require evaluation of the intersection formula; monomers fully inside or outside a bin contribute weights of 1 or 0, respectively. The special case $r_0 = 0$ uses the fully-inside condition, avoiding division by zero.

To eliminate geometric artifacts arising from unequal shell volumes, bins are constructed to have equal volumes. For $N_B$ bins spanning capsid radius $R_{\text{cap}}$, each shell has volume:
\begin{equation}
    V_{\text{shell}} = \frac{4\pi R_{\text{cap}}^3}{3 N_B}
\end{equation}
Requiring equal volume for each shell gives bin boundaries:
\begin{equation}
    R_i = R_{\text{cap}} \left(\frac{i}{N_B}\right)^{1/3}
\end{equation}
ensuring uniform shell volumes across all bins. Density profiles were normalized by bin volume and averaged over all snapshots in the equilibrated regime.

\subsection{Angular Monomer Density Profile Calculation}
\label{sec:SI_AngularProfile}

Taking the capsid center as the origin, each charge site $s$ at position $\vec{C}_s$ defines a radial unit axis $\hat{u}_s = \vec{C}_s / \lvert\vec{C}_s\rvert$. For a monomer centered at $\vec{r}_i$, the deviation angle from this axis is 
\begin{equation}
    \theta_{is}=\arccos\!\left(\frac{\vec{r}_{i}\cdot\hat{u}_{s}}
    {\lvert\vec{r}_{i}\rvert}\right).
\end{equation}
For each analyzed frame, this angle was evaluated separately for every monomer--site pair. A pair was assigned to the cone around the site $s$ when the following is satisfied.
\begin{equation}
    \theta_{is}  < \theta_{c}
\end{equation}
where $\theta_c$ is the set boundary angle from the site axis, set at 9 degrees with 45 bins across all the angular analysis present in the main-text. Only pairs passed the filtering criterion were retained for angular binning. For the 2-bead N-terminal representation, one axis was defined per N-terminal using the based bead. The mobile tail was not treated as a separate axis. Monomer centers within the radial range 0--12~nm were included without partial bead-volume weighting.

The retained angles, $0\leq \theta_{is}<\theta_c$, were divided into $N_\theta$ equal-volume angular bins within the cone. In spherical coordinates, the volume of the bin $k$, spanning $[\theta_k, \theta_{k+1})$, is obtained by integrating the volume element $r^2\sin\theta\,dr\,d\theta\,d\phi$:
\begin{equation}
\begin{aligned}
    V_k
    &=\int_{R_{\rm lo}}^{R_{\rm hi}}\int_{0}^{2\pi}
      \int_{\theta_k}^{\theta_{k+1}}r^2\sin\theta\,d\theta\,d\phi\,dr \\
    &=\frac{2\pi}{3}\left(R_{\rm hi}^{3}-R_{\rm lo}^{3}\right)
      \left(\cos\theta_k-\cos\theta_{k+1}\right).
\end{aligned}
\end{equation}
Because the radial limits are common to all bins (taken at $R_{\rm hi}=12$, $R_{\rm lo}=0$), equal bin volumes are obtained by spacing the edges uniformly in $\cos{\theta}$:
\begin{equation}
    \cos\theta_k=1-\frac{k}{N_\theta}\left(1-\cos\theta_c\right),
    \qquad k=0,\ldots,N_\theta.
\end{equation}
Thus, every bin has volume
\begin{equation}
    V_k=\frac{2\pi}{3}\left(R_{\rm hi}^{3}-R_{\rm lo}^{3}\right)
    \frac{1-\cos\theta_c}{N_\theta},
\end{equation}
and its angular monomer density is
\begin{equation}
    \rho_k = \frac{H_k}{N_{\rm site}N_{\rm frame}V_k}
\end{equation}
where $H_k$ is the accumulated count of the accepted monomer--site pairs in bin $k$, $N_{\rm site}$ is the number of site axes, and $N_{\rm frame}$ is the number of analyzed frames. The calculated profile was mirrored about $\theta=0$ only for displayed.

\subsection{Time-series Estimate of Persistence-Length Uncertainty}
\label{sec:SI_ErrorAnalysis}

For each ion condition, the uncertainty assigned to $L_p$ was estimated from the temporal fluctuation within its analyzed trajectory. The central values shown in main-text Fig.~\ref{fig:Lp_vs_ion} was retained from the persistence-length fit to the bond-vector correlation function averaged over the analyzed interval, following the procedure described in Sec.~\ref{sec:SI_method_TrjAnalysis}. To characterize its temporal fluctuations, the same fitting procedure was also applied separately to every saved frame, obtaining the per-frame estimates $L_{p,k}$ and used in subsequent statistical analysis.

Let $\sigma_{L_p}$ be the standard deviation of the resulting $L_{p,k}$ series. The uncertainty shown in Fig.~\ref{fig:Lp_vs_ion} was calculated as the following,
\begin{equation}
    \delta L_p=\frac{\sigma_{L_p}}{\sqrt{N_s}},
    \qquad
    N_s=\frac{T}{\tau},
\end{equation}
where $T$ is the length of analyzed trajectory in nanoseconds, and $\tau$ is the correlation time for each trajectory, obtained separately for from an exponential fit to the $R_g$ autocorrelation for each salt condition.

\subsection{Realized Salt Density Correction}
\label{sec:SI_IonCorrection}

Ion numbers were assigned as $N = cV_{\rm cap}$, where $c$ is the nominal salt concentration (10, 60, 100 mM) and $V_{\rm cap} = \frac{4}{3}\pi R_{\rm cap}^3 \approx 7238.23$~nm$^3$ is the geometric interior volume of the capsid at its inner radius $R_{\rm cap}=12$~nm. Because the solvent does not occupy this entire volume, the realized concentration inside the capsid is higher than the nominal label.

The realized salt concentrations ($c_{\rm real}$) are determined by $c_{\rm real} = c\,(N_{\rm sys} / N_{\rm w})$, where $N_{\rm w}$ is the number of water molecules present in the system, and $N_{\rm sys}$ is the total number of beads occupying the volume, including ionic solvent, polymer, and the protruded sites. In the present model, all aforementioned species interact with the coarse-grained water molecules with a Lennard-Jones diameter of $\sigma=0.47$~nm (see Table~\ref{tab:Supp_InteractionTab}). Every species therefore occupies the same effective volume. As a result, the calculated realized salt concentrations in the 10, 60, and 100~mM salt series are 10.3, 62.2, and 104.5~mM, a maximum offset compared to the nominal concentration of approximately 4.5\% within the three salt conditions. The correction factor, defined as $N_{\rm sys} / N_{\rm w}$, agrees to within 1.7\%. Among the capsid-charge conditions in Fig.~\ref{fig:DensityProfile_Addition}, the correction factor varies by approximately 2.7\%. Among the N-terminal representations in Fig.~\ref{fig:DensityProfile_60mM_Addition}, it varies by approximately 0.9\%. Each of these variations within the comparison conditions is more than an order of magnitude lower than the 67\% increase from 60 to 100~mM that drives the salt-dependent changes in the monomer density profiles. Therefore the corrections to nominal salt concentrations do not alter any comparison of conditions reported in the main-text.

\subsection{Computational Cost}
\label{sec:SI_ComputationalCost}

The capsid system contained approximately $1.8 \times 10^5$ coarse-grained particles, with the majority being polarizable solvent beads. Simulations on AMD EPYC 9654 processors advanced at approximately 23~ns/day, those on AMD EPYC 9655 processors advanced at approximately 28~ns/day, and those on AMD EPYC 7532 processors at approximately 17~ns/day. The aggregate cost of the capsid production simulations reported here was approximately $1.5\times 10^6$ core-hours.

PPPM electrostatics accounted for approximately 60\% of total runtime, making system volume, rather than genome length, the dominant cost driver, as the genome constitutes only $<1$\% of total particles. To estimate the cost of larger solvent-filled system, we assumed fixed solvent density and proportional scaling of the simulation box. The extrapolation is based on the current system with T=3 capsid (inner radius of $\sim$12~nm \cite{Cadena-Nava2012, Speir1995CCMV}) and particle count of approximately $1.8\times 10^5$. At fixed solvent density, the number of particles grows as $R^3$, and the long-range electrostatics (PPPM) scale as $N~{\rm log}(N)$. Thus we estimate that, for a system with T=4 dimension (inner radius at around 13~nm \cite{Wynne1999, Olson1990NBV}) would cost about $1.3\times$ more than the present, and for a system with T=7 dimension (inner radius at around 18 nm \cite{Saper2013}), it costs approximately $3.7\times$ more than the present.

\section{Supplemental Figures}
\label{sec:SI_figures}

\begin{figure}[H]
    \centering
    \includegraphics[width=0.95\linewidth]{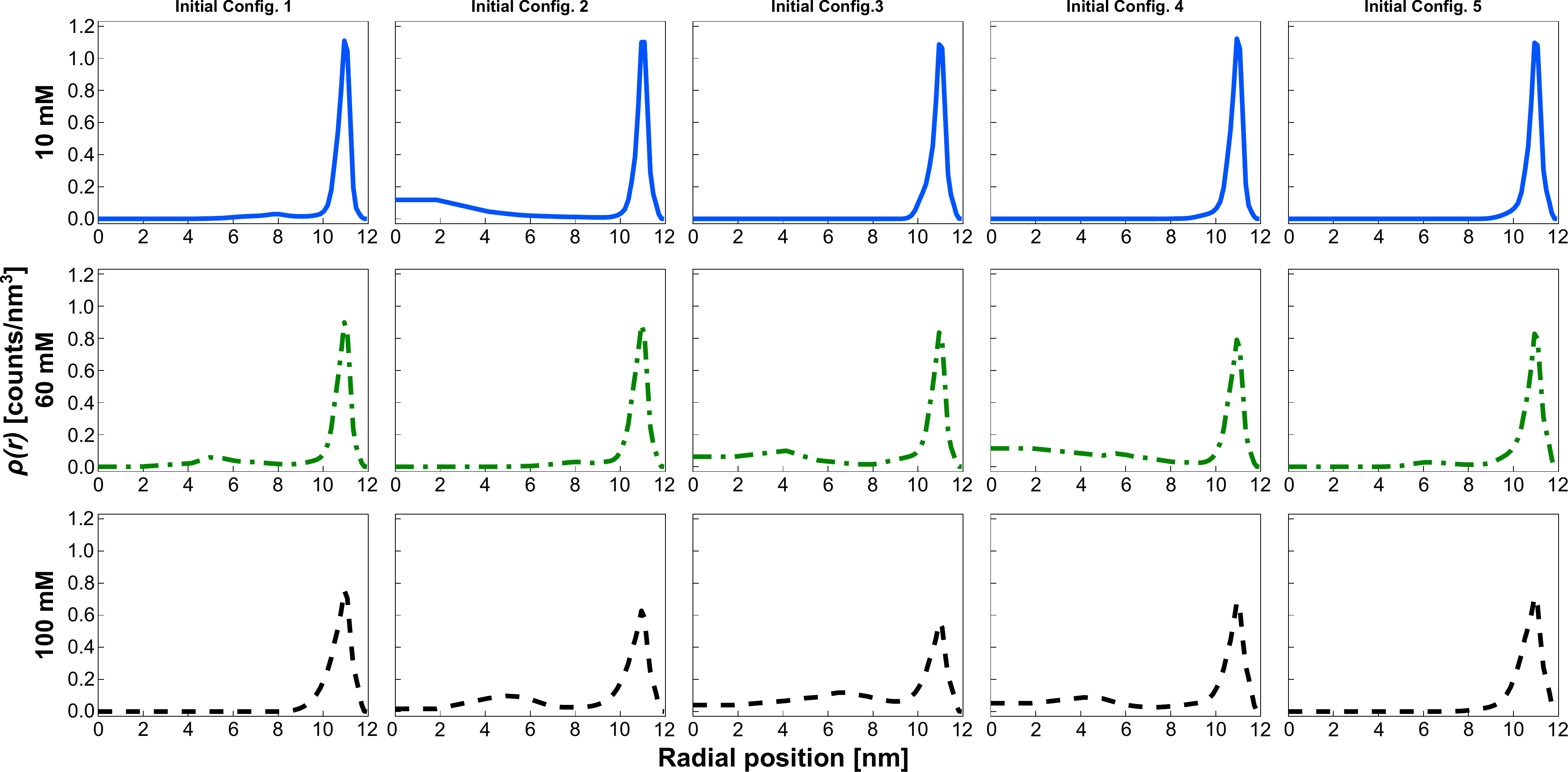}
    \caption{\textbf{Radial monomer density profiles $\rho(r)$ for all five initial configurations at three salt concentrations.} For each row, from left to right, plots correspond to initial configurations 1--5 (Fig.~\ref{fig:Vis_DiffConfig}a--e), respectively. All simulations use 180 N-terminals and overcharging ratio 1.6. Curves show 10~mM (blue solid), 60~mM (green dot-dashed), and 100~mM (black dashed).}
    \label{fig:Individual_Profiles_Standard}
\end{figure}

\begin{figure}[H]
    \centering
    \includegraphics[width=0.4\linewidth]{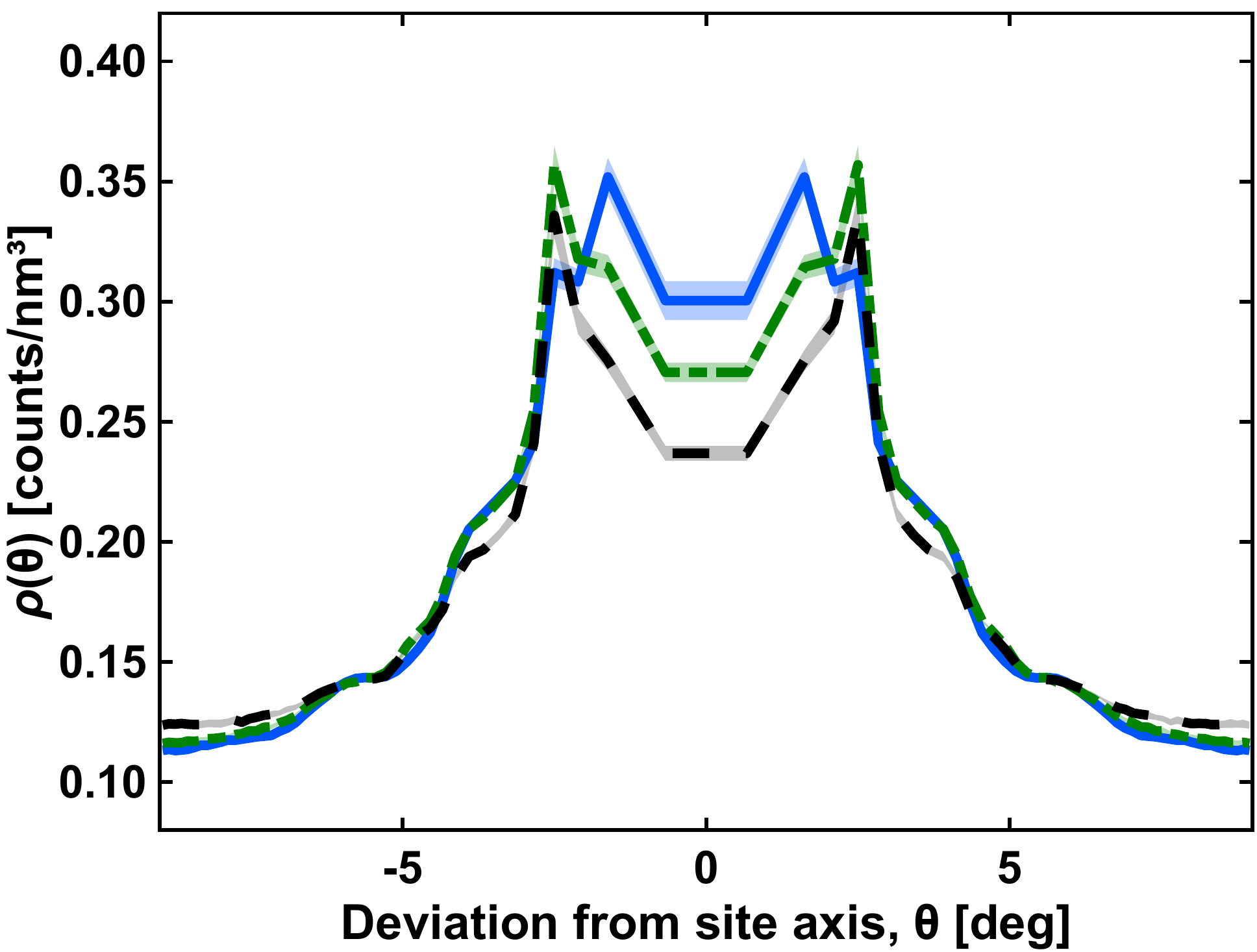}
    \caption{\textbf{Mean angular monomer density $\rho(\theta)$ as a function of angular deviation from the site-axis at varies of salt concentrations:}  10~mM (blue solid), 60~mM (green dot-dashed), and 100~mM (black dashed). Shaded bands: standard error of mean across the 5 individual profiles}
    \label{fig:Figure_4_AngularProfile}
\end{figure}

\begin{figure}[H]
    \centering
    \includegraphics[width=1\linewidth]{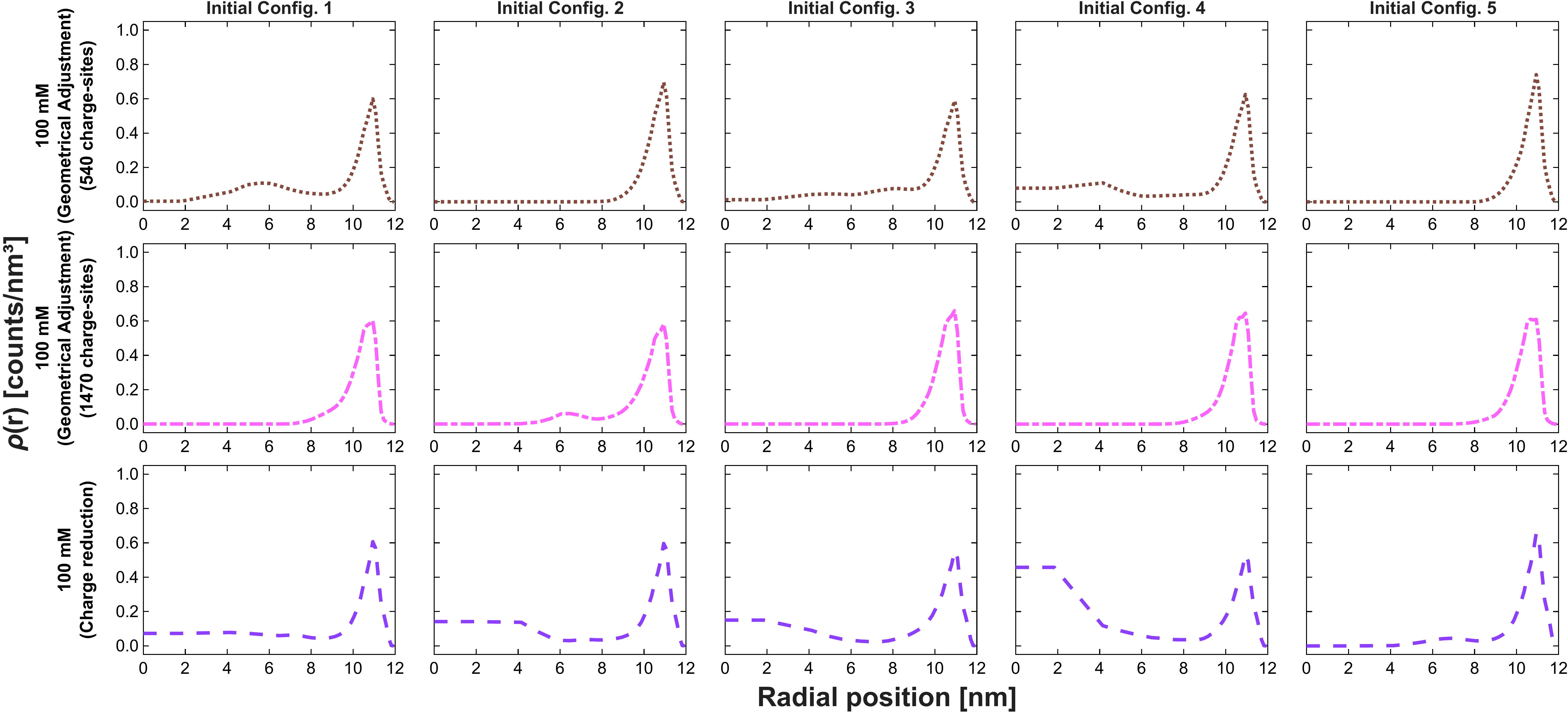}
    \caption{\textbf{Radial monomer density profiles $\rho(r)$ for the geometrical and charge modification cases at 100~mM salt.} For each row, from left to right, plots correspond to initial configurations 1--5 (Fig.~\ref{fig:Vis_DiffConfig}a--e), respectively. Curves show the geometrical adjustment (540 charge sites and 1470 charge sites, overcharging ratio 1.6, bronze dotted and pink dotted-dash) and charge reduction (180 charge sites, overcharging ratio 2.0, purple dashed) cases. Standard case profiles at 100~mM (180 charge sites, overcharging ratio 1.6) are shown in Fig.~\ref{fig:Individual_Profiles_Standard} for comparison.}
    \label{fig:Individual_Profiles_100mMAdditional}
\end{figure}

\begin{figure}[H]
    \centering
    \includegraphics[width=0.9\linewidth]{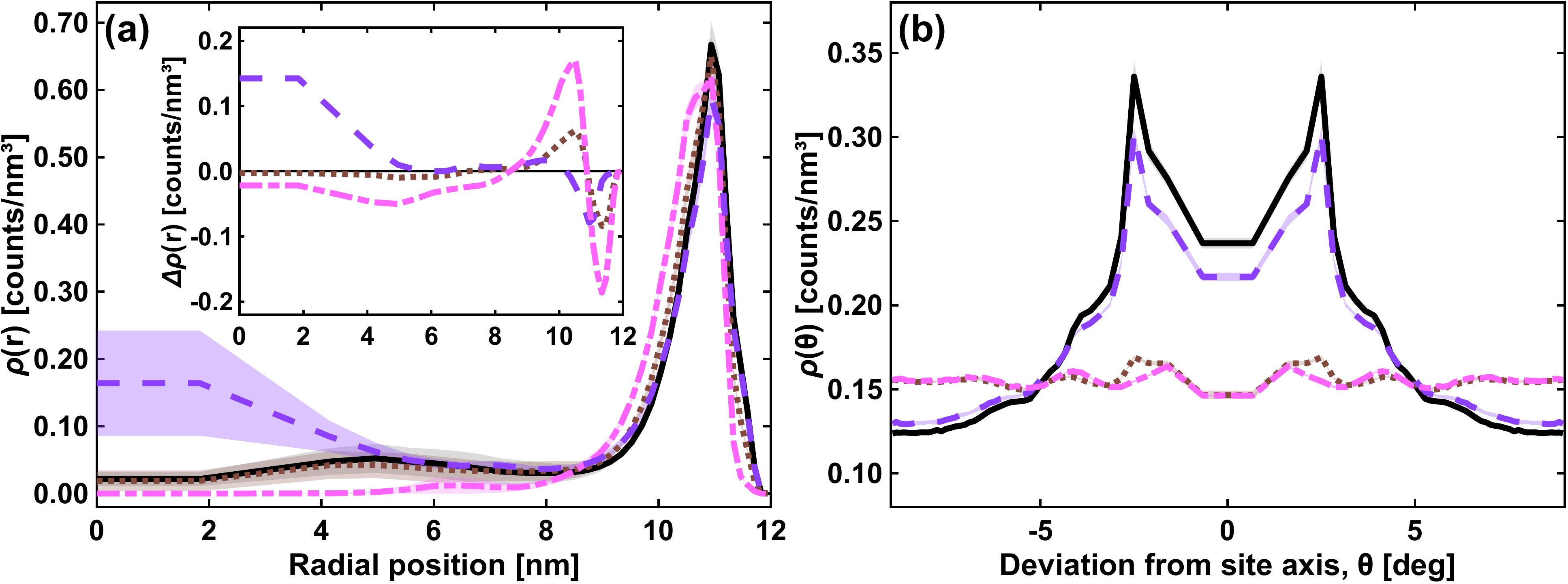}
    \caption{\textbf{Complete comparison of mean radial and angular monomer profiles with different capsid charge conditions at 100 mM of salt concentration.} Mean radial monomer density $\rho(r)$ at 100~mM salt for all four capsid configurations: standard (180 charge sites, overcharging ratio 1.6, black solid), geometrical adjustment (1470 charge sites, overcharging ratio 1.6, pink dot-dashed),geometrical adjustment (540 charge sites, overcharging ratio 1.6, bronze dotted) , and charge reduction (180 sites, overcharging ratio 2.0, purple dashed). Inset: Mean density difference $\Delta\rho(r) = \rho(r) - \rho_{\text{std}}(r)$ relative to the standard profile. (b) Mean angular monomer density profile at 100 mM of salt for standard (black solid), geometrical adjustment (pink dot-dashed for 1470-site and bronze dotted for 540-site), and charge reduction (purple dashed). Shaded bands: standard error of mean across the 5 individual profiles for both panels. }
    \label{fig:Fig_5_4-cases}
\end{figure}

\begin{figure}[H]
    \centering
    \includegraphics[width=1\linewidth]{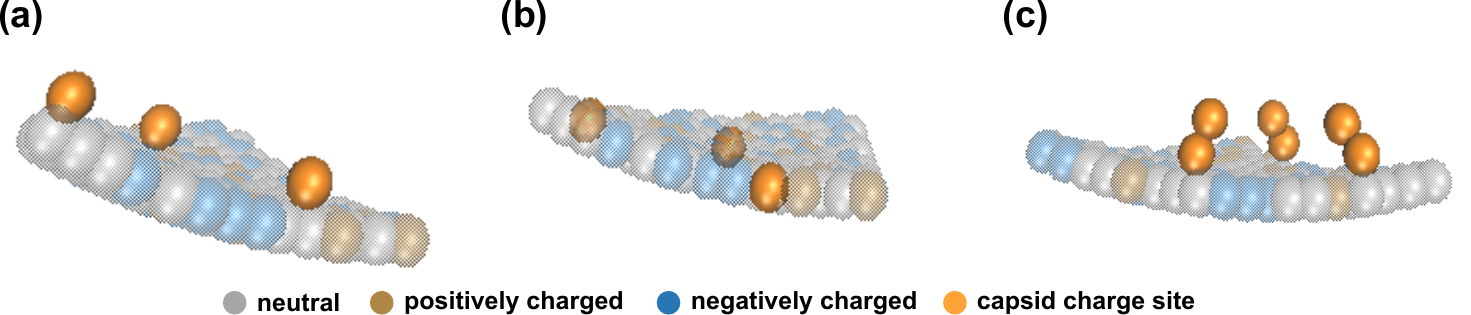}
    \caption{\textbf{Three N-terminal structural representations shown on a portion of the inner capsid.} \textbf{(a)} 1-bead representation: each N-terminal is a single charged sphere protruding into the capsid interior. This is the standard representation used in all simulations unless otherwise stated. \textbf{(b)} 0-bead representation: N-terminal charge is incorporated directly into the inner shell, creating a smooth interior surface. \textbf{(c)} 2-bead representation: each N-terminal is modeled as a flexible two-bead chain connected by a FENE bond. In all cases, 180 discrete charge sites and overcharging ratio 1.6 are maintained. Silver, brown, and blue beads represent neutral, positively charged, and negatively charged inner shell beads, respectively; orange beads represent N-terminal charge sites. Each charge site carries $+1.2986e$ in the 1-bead and 0-bead representations, and $+0.6493e$ per bead in the 2-bead representation, yielding the same total capsid charge in all cases.}
    \label{fig:Vis_NterminalRep}
\end{figure}

\begin{figure}[H]
    \centering
    \includegraphics[width=1\linewidth]{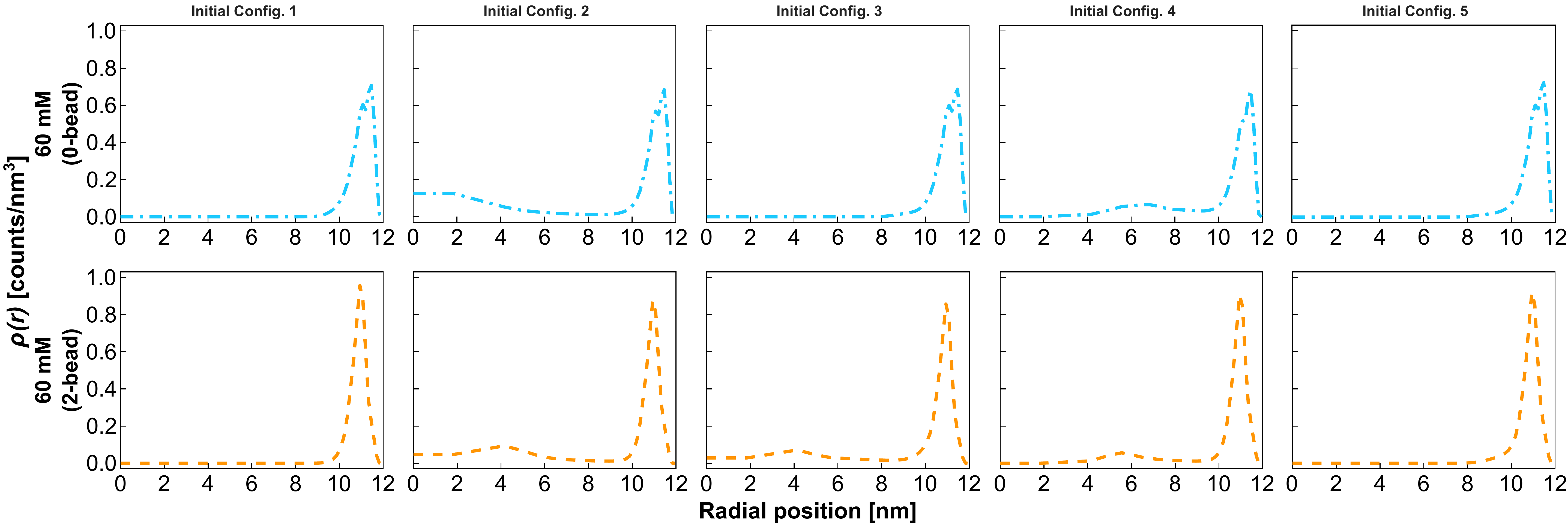}
    \caption{\textbf{Radial monomer density profiles $\rho(r)$ for three N-terminal representations at 60~mM salt.} For each row, from left to right, plots correspond to initial configurations 1--5 (Fig.~\ref{fig:Vis_DiffConfig}a--e), respectively. The three N-terminal representations are illustrated in Fig.~\ref{fig:Vis_NterminalRep}. Curves show the 0-bead (cyan dot-dashed) and 2-bead (orange dashed) representations. Standard 1-bead profiles at 60~mM are shown in Fig.~\ref{fig:Individual_Profiles_Standard} for comparison.}
    \label{fig:Individual_Profiles_60mMAdditional}
\end{figure}

\end{document}